\documentclass[sn-standardnature]{sn-jnl}
\usepackage{xspace}

\jyear{2021}%

\theoremstyle{thmstyleone}%
\theoremstyle{thmstyletwo}%

\theoremstyle{thmstylethree}%

\newcommand{\araa}{Annu. Rev. Astron. Astrophys.}   
\newcommand{\aj}{Astron. J.}   
\newcommand{\apj}{Astrophys. J.}   
\newcommand{\apjl}{Astrophys. J. Lett.}   
\newcommand{\apjs}{Astrophys. J. Suppl. Ser.}   
\newcommand{\apss}{Astrophys. Space Sci.}   
\newcommand{\aap}{Astron. Astrophys.}   
\newcommand{\icarus}{Icarus}   
\newcommand{\jqsrt}{J. Quant. Spectrosc. Radiat. Transf.} 
\newcommand{\mnras}{Mon. Not. R. Astron. Soc.}   
\newcommand{\planss}{Planet. Space Sci.}   
\newcommand{\pasp}{Publ. Astron. Soc. Pac.}   

\begin{document}

\title[JWST detection of abundant hydrocarbons in a buried nucleus]{Abundant hydrocarbons in a buried galactic nucleus with signs of carbonaceous grain and polycyclic aromatic hydrocarbon processing}


\author*[1]{\sur{Garc\'ia-Bernete}, \fnm{I.}}\email{igbernete@cab.inta-csic.es}
\author[2]{\sur{Pereira-Santaella}, \fnm{M.}}
\author[3]{\sur{Gonz\'alez-Alfonso}, \fnm{E.}} 
\author[2]{\sur{Ag\'undez}, \fnm{M.}}
\author[4,5]{\sur{Rigopoulou}, \fnm{D.}}
\author[4]{\sur{Donnan}, \fnm{F.\,R.}}
\author[2]{\sur{Speranza}, \fnm{G.}}
\author[4]{\sur{Thatte}, \fnm{N.}}

\affil*[1]{\orgdiv{Centro de Astrobiolog\'ia (CAB)},\orgname{CSIC-INTA}, \orgaddress{\street{Camino Bajo del Castillo s/n}, \city{Villanueva de la Ca\~nada, Madrid}, \postcode{E-28692}, \country{Spain}}}
\affil[2]{\orgdiv{Instituto de F\'isica Fundamental}, \orgname{CSIC}, \orgaddress{\street{Calle Serrano 123}, \city{Madrid}, \postcode{E-28006}, \country{Spain}}}
\affil[3]{\orgdiv{Departamento de F\'isica y Matem\'aticas}, \orgname{Universidad de Alcal\'a}, \orgaddress{\street{Campus Universitario}, \city{Alcal\'a de Henares, Madrid}, \postcode{E-28871}, \country{Spain}}}
\affil[4]{\orgdiv{Department of Physics}, \orgname{University of Oxford}, \orgaddress{\street{Keble Road}, \city{Oxford}, \postcode{OX1 3RH}, \country{UK}}}
\affil[5]{\orgdiv{School of Sciences}, \orgname{European University Cyprus}, \orgaddress{\street{Diogenes street}, \city{Engomi}, \postcode{1516 Nicosia}, \country{Cyprus}}}


\abstract{Hydrocarbons play a key role in shaping the chemistry of the interstellar medium (ISM), but their enrichment and relationship with carbonaceous grains and polycyclic aromatic hydrocarbons (PAHs) still lack clear observational constraints. We report JWST NIRSpec+MIRI/MRS infrared (IR; $\sim$\,3-28\,$\mu$m) observations of the local ultra-luminous IR galaxy (ULIRG) IRAS\,07251$-$0248, revealing the extragalactic detection of small gas-phase hydrocarbons such as benzene (C$_6$H$_6$), triacetylene (C$_6$H$_2$), diacetylene (C$_4$H$_2$), acetylene (C$_2$H$_2$), methane (CH$_4$), and methyl radical (CH$_3$) as well as deep amorphous C–H absorptions in the solid phase. The unexpectedly high abundance of these molecules indicates an extremely rich hydrocarbon chemistry, not explained by high-temperature gas-phase chemistry, ice desorption or oxygen depletion. Instead, the most plausible explanation is the erosion and fragmentation of carbonaceous grains and PAHs. This scenario is supported by the correlation between the abundance of one of their main fragmentation products, C$_2$H$_2$, and cosmic ray (CR) ionization rate for a sample of local ULIRGs. These hydrocarbons are outflowing at $\sim$160\,km/s, which may represent a potential formation pathway for hydrogenated amorphous grains. Our results suggest that IRAS\,07251$-$0248 might not be unique but represents an extreme example of the commonly rich hydrocarbon chemistry prevalent in deeply obscured galactic nuclei.}

\maketitle
\vspace{-0.5truecm}

Carbonaceous dust evolves within the interstellar medium (ISM), being processed from its formation, in circumstellar outflows, to its destruction, and potential reformation in dense regions \cite{Jones11}. The evolution of carbonaceous dust, including polycyclic aromatic hydrocarbons (PAHs) and carbonaceus grains is likely connected to the diversity of dust properties across different sources. We refer to carbonaceous grains as solid-phase dust particles rich in carbon, including hydrogenated amorphous carbon. Hereafter we will use the term carbonaceous grains and hydrogenated amorphous carbon grains (HAC or a-C:H) interchangeably.

PAHs belong to the lower end of the ISM dust distribution size and account for up to $\sim$10\% of interstellar carbon \cite{Tielens2021_book}. These molecules are excited by UV/optical photons, producing infrared (IR) features that can contribute significantly to the total IR emission \cite{Smith07a}. PAH features are frequently used to estimate star formation rates in galaxies and to probe the different physical conditions in the ISM (see e.g. \cite{Tielens2021_book}).

Different processes can drive the destruction of carbonaceous grains and PAH molecules, including photodissociation, ionization, cosmic rays, and shocks \cite{Micelotta10,Pino19}. Fragmentation and erosion of carbonaceous grains \cite{Alata15,Murga23} and PAHs \cite{Monfredini19} may inject small hydrocarbons and carbons chains into the gas-phase, enriching the ISM with organic molecules. Furthermore, when small hydrocarbons are abundant in the gas phase, they can react with existing PAHs, generating bottom-up formation pathways that lead to the growth of larger PAH molecules (e.g. \cite{Goicoechea25} and references therein). For instance, alkanes (e.g. CH$_4$), alkynes (e.g. C$_2$H$_2$, C$_4$H$_2$, C$_6$H$_2$), and methyl groups (e.g. CH$_3$) might be part of larger PAH molecules, which are mainly composed of benzene-like rings (C$_6$H$_6$). The structure of benzene (C$_6$H$_6$) serves as the fundamental example to define aromaticity. Cyclic hydrocarbons alternate single and double bonds, with each carbon attached to one hydrogen and two other carbons. This simple aromatic hydrocarbon is a key building block for PAHs. 

Previous IR observations, from Galactic to extragalactic environments, have revealed the presence of C$_2$H$_2$ and/or HCN \cite{Carr95,vanDishoeck98,Chiar00,Lahuis07}. Recent JWST observations have also shown that protoplanetary disks and Galactic star-forming regions are rich in organic gas-phase molecules and ices (e.g. \cite{McClure23, Tabone23, Rocha24}). Most studies modeling these bands indicate a high hydrocarbon abundance, with a carbon-to-oxygen (C/O) ratio above 0.8. 

Strong ice features are also observed in deeply buried nuclei of ultraluminous infrared galaxies (U/LIRGs; \cite{Spoon22}). The large columns of gas and dust in their nuclear region usually prelude the detection of the powering sources in the optical and X-rays \cite{Bernete22b}. The most extremes cases are known as compact obscured nuclei (CONs), where a combination of a compact starburst (SB) and AGN activity might power the IR emission but the dominant power source is still under debate (e.g. \cite{Aalto15,Pereira-Santaella21,Bernete24b}). While molecular bands such as CO, CO$_2$, H$_2$O, C$_2$H$_2$ and HCN are typically found in galaxies \cite{Lahuis07, Gonzalez-Alfonso24, Pereira24, Bernete24b,Buiten24, Buiten25}, the detection of many small gas-phase hydrocarbons in extragalactic objects has so far remained elusive. 

Here we focus on the JWST NIRSpec+MIRI/MRS spectrum (3-28\,$\mu$m) of the eastern nucleus of IRAS\,07251$-$0248 due to its exceptionally rich spectral features. Hereafter, we will refer to the eastern nucleus simply as IRAS\,07251$-$0248. This nucleus is one of the most dust-obscured in the local Universe (z$<$0.3; e.g. \cite{Bernete22b}) making IRAS\,07251$-$0248 an ideal target for studying the properties of the ISM. In particular, it allows for the investigation of the numerous molecular absorptions expected in its highly gas- and dust-rich environment. The source exhibits signatures typical of a buried nuclei such as a deep OH\,65\,$\mu$m absorption \cite{Gonzalez-Alfonso15} or a compact sub-mm core (r$<$20\,pc; {\textcolor{blue}{Gonz\'alez-Alfonso et al. in prep.}}, \cite{Pereira-Santaella21}). 
The JWST observations reveal that various hydrocarbons, including gas-phase C$_6$H$_6$, C$_6$H$_2$, C$_4$H$_2$, C$_2$H$_2$, CH$_4$ and CH$_3$ are present and are abundant in this extragalactic environment. The solid phase carbonaceous grains also show deep absorption features against the strong mid-IR continuum.\\ 

\begin{figure*}
\centering
\par{
\includegraphics[width=10.cm, clip, trim=10 0 10 0]{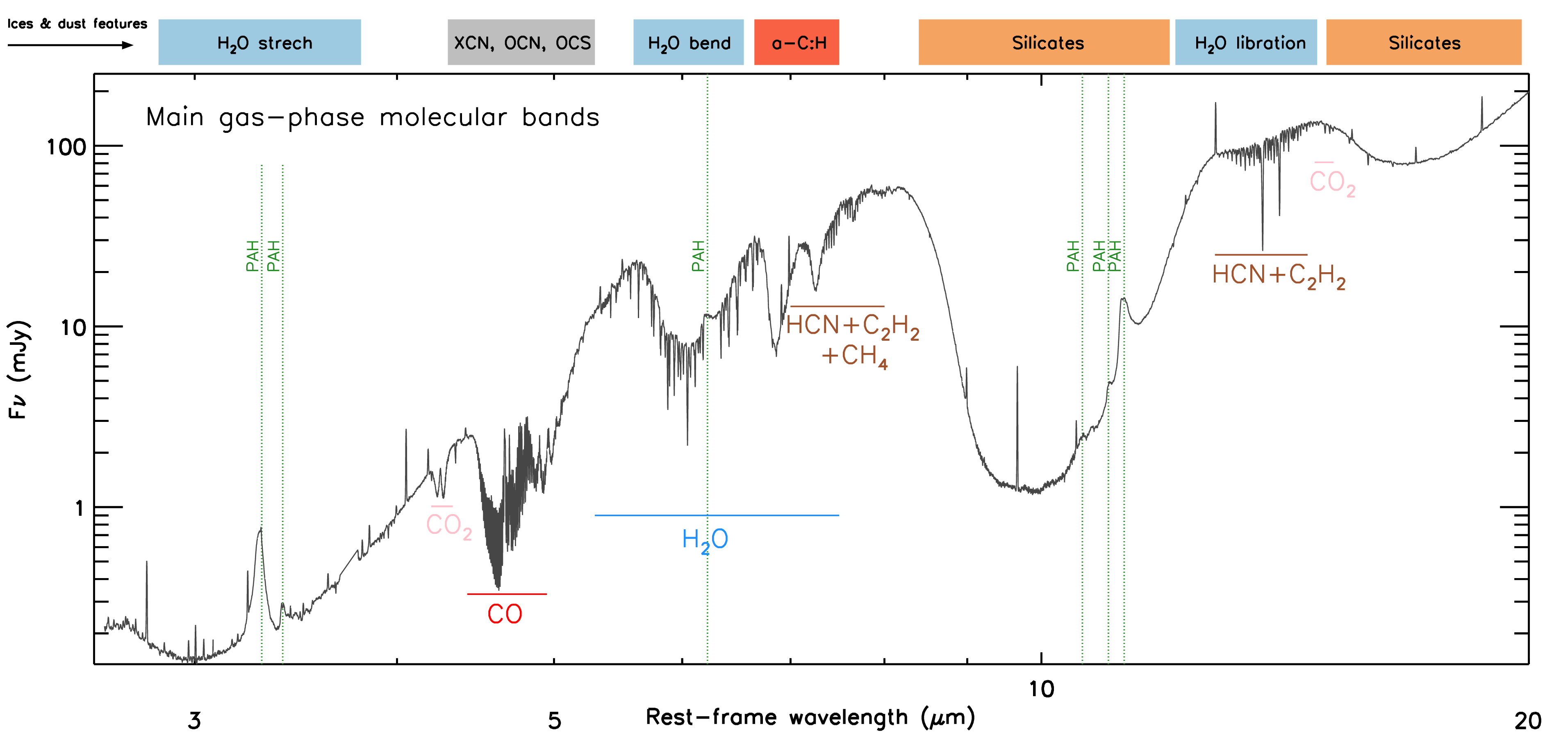}
\par}
\caption{JWST near- and mid-IR spectrum of the eastern nucleus of IRAS\,07251$-$0248. The spectrum was extracted assuming it is a point source ($\sim$0.4'' aperture diameter at 10\,$\mu$m; see Methods for further details). The main molecular gas-phase bands detected in this source are indicated: CO, CO$_2$, H$_2$O, HCN, CH$_4$ and C$_2$H$_2$. The vertical green dotted lines correspond to the clearly detected PAH bands in IRAS\,07251$-$0248. The top shaded regions represent the typical extent of the ices and dust features.}
\label{restframe_spec}
\end{figure*}

{\bf{Gas-phase molecular bands: }} 
The molecular gas-phase absorption bands typically found in galaxies are strong and clearly detected in IRAS\,07251$-$0248 (see Fig. \ref{restframe_spec}). 
These include the fundamental bands of $^{12}$CO at $\sim$4.45--4.95~$\mu$m, $^{13}$CO at $\sim$4.8$\mu$m, H$_2$O $v_2$ at $\sim$5.3-7.2~$\mu$m, C$_2$H$_2$ $v_5$ at 13.7~$\mu$m, HCN $v_2$ at 14.0~$\mu$m, and the CO$_2$ $v_3$ and $v_2$ bands at $\sim$4.25 and 15.0~$\mu$m, respectively (see e.g. \cite{Lahuis07}). 
This object presents some less common absorptions in the $\sim$7.1--7.9$\mu$m range that correspond to the HCN 2$v_2$ overtone ($\sim$7.1$\mu$m), the C$_2$H$_2$ $v_4$+$v_5$ combination band ($\sim$7.5$\mu$m), and the CH$_4$ $v_4$ fundamental band at $\sim$7.7$\mu$m (see e.g. \cite{Buiten25}). In addition, other mid-IR absorption bands, which have not been previously detected in this kind of extragalactic environments, are observed, corresponding to C$_6$H$_6$ $v_4$ at 14.8$\mu$m, C$_4$H$_2$ $v_8$ at 15.9$\mu$m, C$_6$H$_2$ $v_{11}$ at 16.1$\mu$m, and CH$_3$ $v_2$ at 16.5$\mu$m.

\begin{figure*}
\centering
\par{
\includegraphics[width=12.6cm, clip, trim=10 10 10 0]{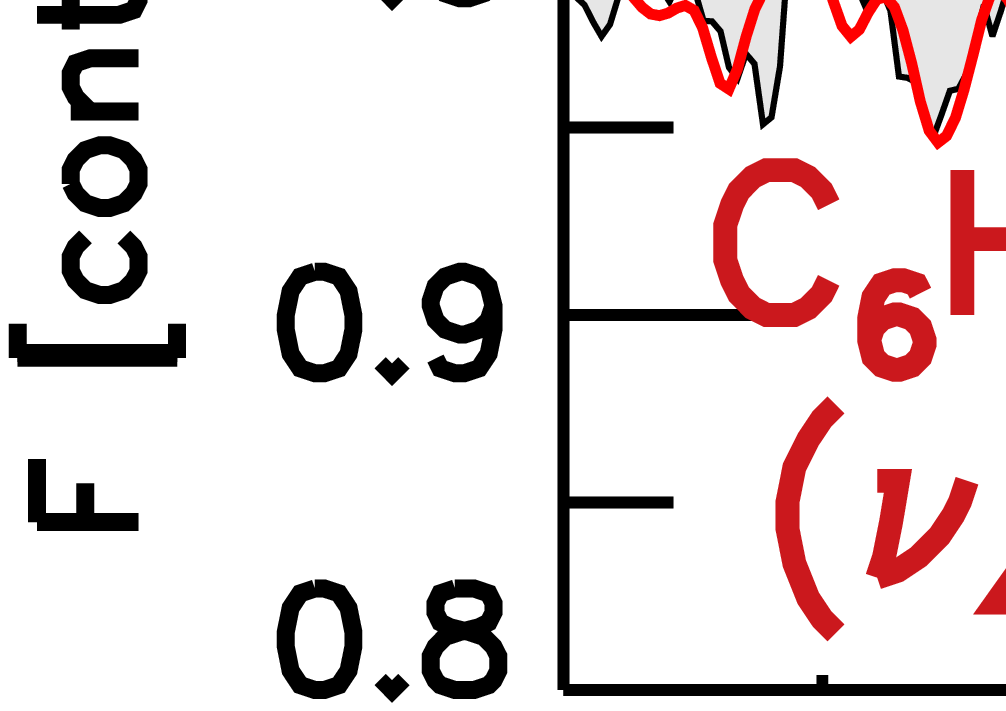}
\par}
\caption{Main mid-IR gas-phase molecular rovibrational bands in IRAS\,07251$-$0248. The top panel shows the $\sim$7-8\,$\mu$m molecular bands and the middle and bottom panels the $\sim$13-18\,$\mu$m bands. Each panel shows the {\textit{JWST}}/MIRI-MRS rest-frame continuum-normalized spectra (Extended data Fig. 1; black solid line filled in gray) together with the total best-fit model (red solid line). The contribution of the fit by the different species are shown with an offset ($+$0.15) to improve clarity: C$_6$H$_6$ (dark red), C$_6$H$_2$ (blue), C$_4$H$_2$ (deep pink), CH$_4$ (magenta), CH$_3$ (green), C$_2$H$_2$ (brown), CO$_2$ (light pink) and HCN (orange).}
\label{model}
\end{figure*}

Since these bands are detected in absorption, we determined the column densities and rotational temperatures of the ground states of these molecules using LTE radiative transfer models. 
For the bands of C$_2$H$_2$, CH$_4$, HCN, and H$_2$O in the 5-8 and 13-15\,$\mu$m ranges, which are heavily blended, we created a common model following the methodology presented in \cite{Alfonso14,Gonzalez-Alfonso24} (see Methods). In the 14-18\,$\mu$m range, the observed bands, including some of the high-J transitions of the C$_2$H$_2$ and HCN, are well separated and we modeled them individually (Fig. \ref{model} and Extended Data Fig. 1). The fitted H$_2$O rovibrational band is shown in Extended Data Fig. 2. In addition to the continuum responsible for molecular absorption, an extra mid-IR component may dilute the bands, which we model with a covering factor (f$_{\rm cov}$; see Methods).

HCN, C$_2$H$_2$ and CH$_4$ have been previously observed in ULIRGs \cite{Lahuis07, Alonso-Herrero24, Buiten25}. C$_6$H$_6$, C$_6$H$_2$, C$_4$H$_2$, and CH$_3$ have been detected in the Galactic ISM \cite{Feuchtgruber00, Cernicharo01, Garcia-Hernandez16}, while C$_6$H$_6$, C$_6$H$_2$, and C$_4$H$_2$ have also been identified in the extragalactic planetary nebula SMP LMC 11, an object located in the Large Magellanic Cloud \cite{Bernard-Salas06}. However, their presence and role in extragalactic environments remain largely unexplored. Here, we report the detection of these gas-phase hydrocarbon molecules in an extragalactic environment, specifically in a ULIRG, extending previous findings to more distant and complex systems (Fig. \ref{model}), with the CH$_3$ radical, expected to be at the root of the HCN formation \cite{Sternberg95}, representing the first extragalactic detection of CH$_3$.

All the gas-phase molecules studied in this work are found to be outflowing ($\sim$160\,km/s; see Methods). For most absorption bands, we find rotational temperatures of $T_{\rm rot}$=150-250\,K (Table~\ref{table_columns}). If the excitation is driven by collision with H$_2$, $T_{\rm rot}$ directly traces the kinetic temperature of the gas ($T_{\rm kin}$) for molecules without permanent dipole moment (all but H$_2$O and HCN in our suite).
Alternatively, if the excitation is dominated by radiation, $T_{\rm rot}$ will tend to the radiation temperature ($T_{\rm rad}$), although for each species the actual $T_{\rm rot}$ will depend on the strength of the bands and the critical densities \cite{Imanishi2016}.

In dense regions (n$_{\rm (H_2)}\gtrsim$10$^5$\,cm$^{-3}$; \cite{Goldsmith01}) gas and dust are close to thermal equilibrium. Therefore, 
regardless of the excitation mechanism, the measured $T_{\rm rot}$ should reflect $T_{\rm kin}$.\\

\begin{table}[ht]
\tiny
\centering
\begin{tabular}{lcccccc}
\hline
Molecule & Vibrational& Wavelength & T$_{\rm rot}$ [K]  &     N [cm$^{-2}$] & Abundance & Covering \\
 &mode & [$\mu$m]  &     & &relative to H& factor (f$_{\rm cov}$) \\
\hline
H$_2$O &$v_2$& 6.2 &185$\pm$25 &(5.4$\pm$1.6)$\times$10$^{18}$& (2.84$\pm_{1.89}^{4.14}$)$\times$10$^{-5}$&0.62$\pm$0.04\\
HCN &2$v_2$  & 7.1& 185$\pm$25 &(2.0$\pm$0.6)$\times$10$^{18}$& (1.07$\pm_{0.71}^{1.57}$)$\times$10$^{-5}$&0.62$\pm$0.04\\
HCN  &$v_2$& 14.0& 100$\pm$25 &(6.6$\pm$2.4)$\times$10$^{17}$& (3.48$\pm_{2.43}^{5.51}$)$\times$10$^{-6}$&0.68$\pm$0.05\\
CH$_4$ &$v_4$ & 7.7  & 185$\pm$25 &(7.9$\pm$2.4)$\times$10$^{17}$& (4.17$\pm_{2.80}^{6.12}$)$\times$10$^{-6}$&0.62$\pm$0.04\\
CH$_3$ &$v_2$ & 16.5 & 220$\pm$20 &(4.5$\pm$0.7)$\times$10$^{16}$& (2.37$\pm_{1.43}^{2.82}$)$\times$10$^{-7}$&0.68$\pm$0.05\\
C$_2$H$_2$ &$v_4$+$v_5$& 7.5& 185$\pm$25 &(1.9$\pm$0.6)$\times$10$^{18}$& (1.00$\pm_{0.68}^{1.49}$)$\times$10$^{-5}$&0.62$\pm$0.04\\
C$_2$H$_2$ &$v_5$ &  13.7& 230$\pm$20 &(5.4$\pm$1.9)$\times$10$^{17}$& (2.83$\pm_{1.98}^{4.44}$)$\times$10$^{-6}$&0.68$\pm$0.05\\%
C$_4$H$_2$ &$v_8$ & 15.9 & 170$\pm$30 &(1.3$\pm$0.2)$\times$10$^{16}$& (7.00$\pm_{4.11}^{8.12}$)$\times$10$^{-8}$&0.68$\pm$0.05\\
C$_6$H$_2$ &$v_{11}$ &16.1 & 150$\pm$50 &(2.7$\pm$1.2)$\times$10$^{14}$& (1.42$\pm_{1.05}^{2.47}$)$\times$10$^{-9}$&0.68$\pm$0.05\\
C$_6$H$_6$ &$v_4$ & 14.8 & 250$\pm$30 &(3.7$\pm$0.8)$\times$10$^{16}$& (1.96$\pm_{1.23}^{2.54}$)$\times$10$^{-7}$&0.68$\pm$0.05\\
CO$_2$ &$v_2$& 15.0& 160$\pm$50 &(3.3$\pm$0.6)$\times$10$^{16}$& (1.65$\pm_{1.07}^{2.15}$)$\times$10$^{-7}$&0.68$\pm$0.05\\
\hline
\end{tabular}                                           
\caption{Derived properties from the molecular bands. We assume that CH$_3$, C$_4$H$_2$, C$_6$H$_2$, C$_6$H$_6$ and CO$_2$ have the same covering factor as that obtained for HCN and C$_2$H$_2$ bands at $\sim$15\,$\mu$m. All uncertainties are quoted at the 1-sigma level.} 
\label{table_columns}
\end{table}

\noindent
{\bf{Dust grain features:}} We detect exceptionally deep silicate, H$_2$O ice and a-C:H absorption features (Fig. \ref{restframe_spec}), which are among the deepest observed in galaxies (see Methods; Extended Data Fig. 3). The observed features indicate a high column density of carbon locked in amorphous carbon dust, while CH$_4$ ice is nearly absent (Extended Data Fig. 4). Detailed fits, column densities, and abundance ratios are provided in Methods.\\

\noindent
{\bf{Fraction of large neutral PAH molecules: }}
The spectrum of IRAS\,07251$-$0248 shows PAH bands from neutral molecules at 3.3 and 11.3\,$\mu$m, while the bands attributed to ionized molecules (i.e., 6.2, 7.7 and 8.6\,$\mu$m features) are heavily diluted by the continuum (Fig. \ref{restframe_spec} and Extended Data Fig. 5). The aliphatic PAH bands at 3.4\,$\mu$m and 3.47\,$\mu$m are also detected (Fig. \ref{restframe_spec} and Extended Data Fig. 5). 

\begin{figure}
\centering
\par{
\includegraphics[width=7.5cm, clip, trim=10 0 10 0]{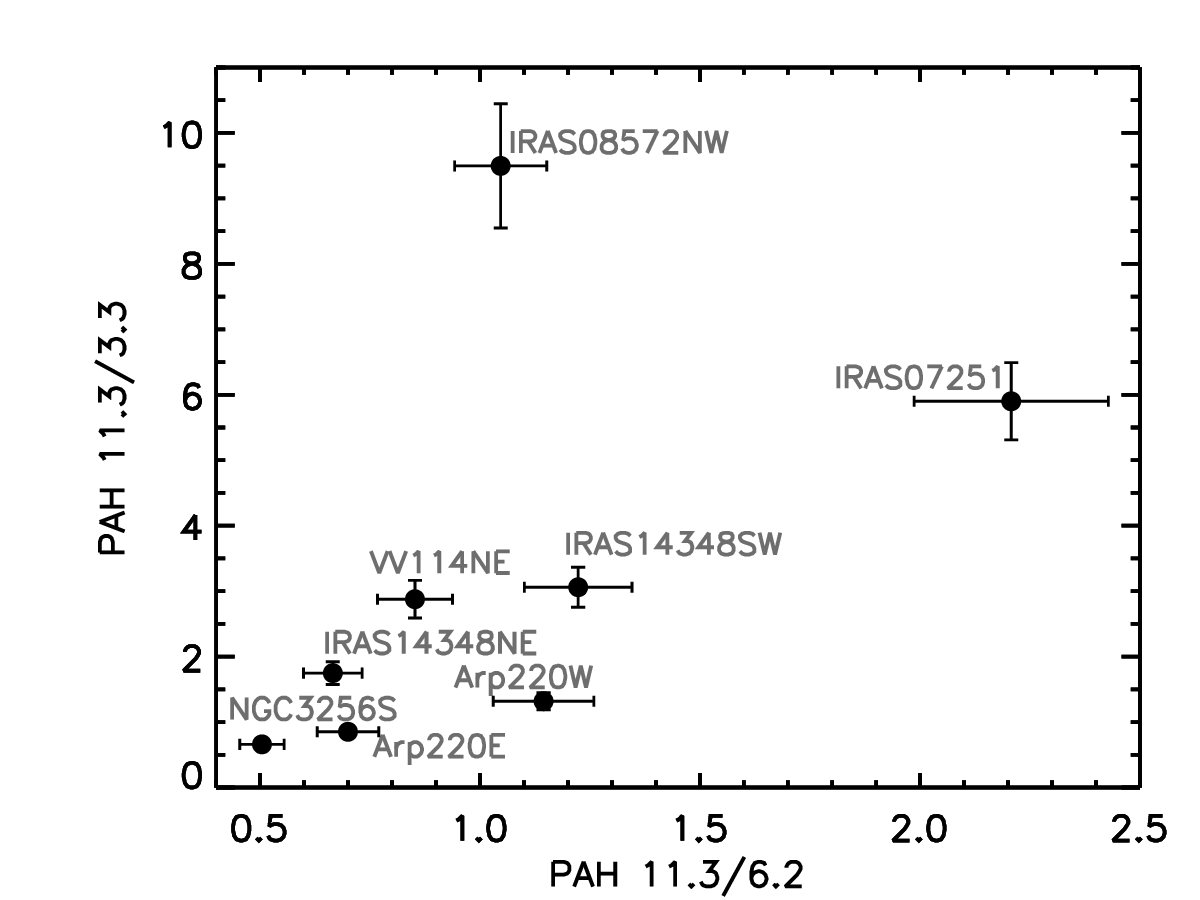}
\par}
\caption{Relationship between key PAH flux ratios in deeply obscured (U)LIRGs.
Comparison of the 11.3/3.3\,$\mu$m and 11.3/6.2\,$\mu$m PAH flux ratios for the sample analyzed in this work. Error bars indicate 1$\sigma$ uncertainties.}
\label{pah_r1}
\end{figure}

We fit the various PAH features, corrected for extinction, by using the method described in \cite{Donnan24} (see Methods). Fig. \ref{pah_r1} shows the 11.3/3.3\,$\mu$m PAH flux ratio, which traces the PAH molecular size \cite{Rigopoulou24}, plotted against the 11.3/6.2\,$\mu$m PAH band flux ratio, which mainly probes the ionization fraction of the PAHs but also depends on their molecular sizes (e.g. \cite{Bernete22d}). 

We find that the 11.3\,$\mu$m PAH feature is enhanced relative to other PAH bands (i.e. 3.3 and 6.2\,$\mu$m) in deeply obscured (U)LIRGs. Furthermore, we find no relation between these PAH ratios and the UV radiation field hardness (see Methods; Extended Data Fig. 6). These results provide evidence for a PAH population dominated by large and neutral molecules. PAH emission is expected to have a spatial distribution more extended than that of the absorption features, which probe the innermost region (Fig. \ref{fig_abs}). Despite the plausible different spatial scales, the anomalous PAH band ratios and strong absorption by gas-phase and solid-phase hydrocarbons and methyl radicals might be related to the remarkable chemistry in the nucleus of this object (see Discussion).

\vspace{5pt}

\noindent
{\bf{Molecular abundances: }} 
Using as a reference the hydrogen nuclei column density (see Methods), we find exceptionally high abundances of gas-phase hydrocarbons, and a-C:H grains in this source (Fig. \ref{abundance_plot}). The gas-phase of these molecules show relatively warm rotational temperatures $\sim$150-250\,K (see Table \ref{table_columns}). Not only hydrocarbons are abundant in IRAS\,07251$-$0248 but also HCN (Table \ref{table_columns}, see Methods). 

\section*{Discussion}\label{sec3}
The spectrum of IRAS\,07251$-$0248 exhibits striking hydrocarbon molecular bands. The derived abundances of gas-phase hydrocarbons (e.g., C$_6$H$_6$, C$_6$H$_2$, C$_4$H$_2$, C$_2$H$_2$, CH$_4$ and CH$_3$) and of carbonaceous grains (solid phase) suggest a hydrocarbon-rich chemistry in this deeply obscured galactic nucleus (See Table \ref{table_columns}). Hereafter, we use the term hydrocarbon-rich environment to describe galactic nuclei where hydrocarbon abundances are similar to or higher than that of H$_2$O. While gas-phase H$_2$O is found at levels comparable to galactic hot cores ($>$10$^{-5}$), organic molecules such as C$_2$H$_2$ and HCN are more abundant ($\sim$10$^{-5}$) than observed in these hot cores ($\sim$10$^{-6}; $\cite{Carr95,van_Dishoeck98,Boonman03}). In the molecular gas of IRAS\,07251$-$0248 probed by JWST, we infer a carbon-to-oxygen (C/O) gas-phase ratio of $\sim$1.03 using the gas-phase molecules listed in Table \ref{table_columns} and the CO column density of $\sim$2.5$\times$10$^{19}\,cm^{-2}$, derived by {\textcolor{blue}{Pereira-Santaella et al., in prep.}}. While contributions from species such as atomic oxygen, O$_2$, and carbon chains are not included, this approach provides a proxy of the C/O gas-phase ratio probed by JWST. We discuss possible scenarios for the origin and enhanced abundance of small hydrocarbons in this source:\\

\begin{figure*}
\centering
\par{
\includegraphics[width=12.4cm, clip, trim=19.5 0 0 0]{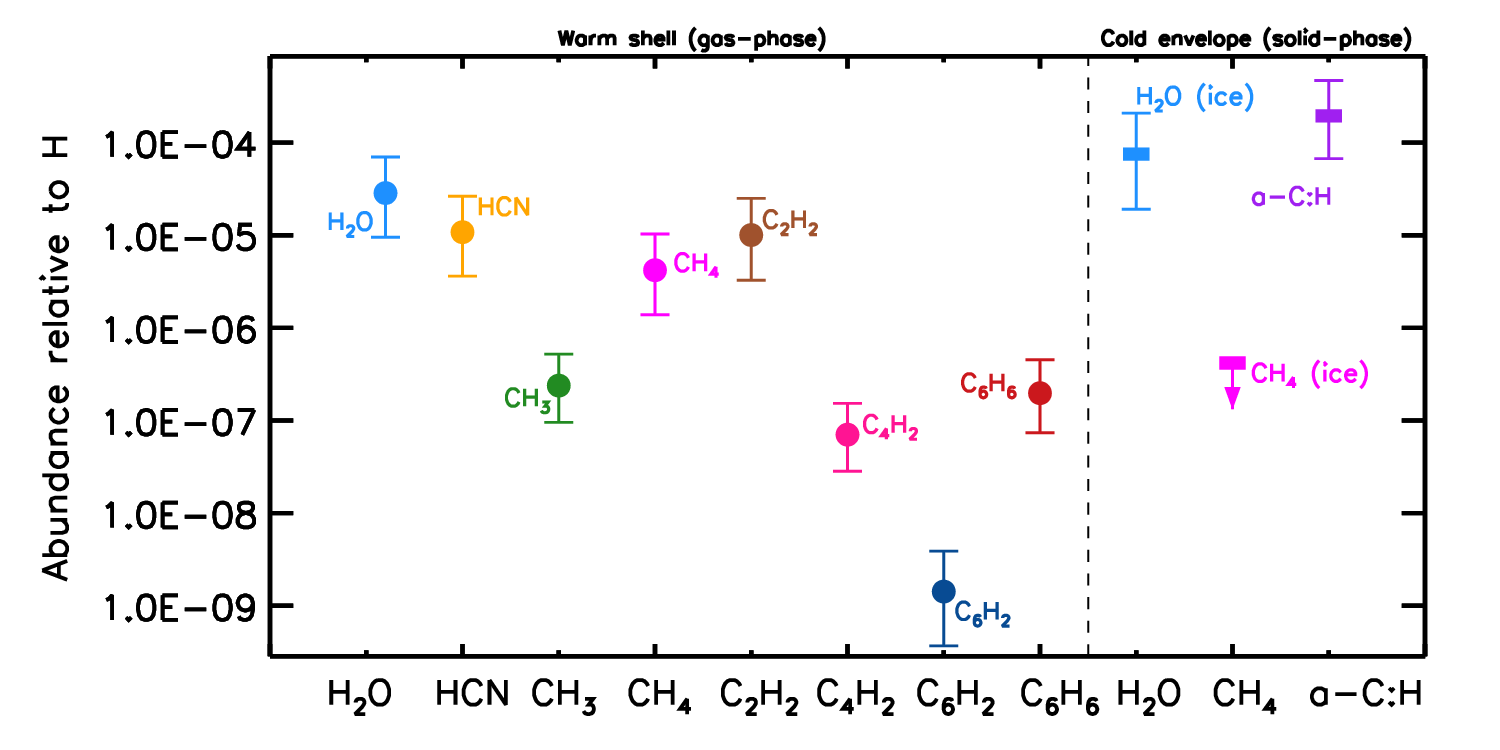}
\par}
\caption{Abundance of organic molecules detected in the nuclear spectrum of IRAS\,07251$-$0248 relative to H. Circles and rectangles represent gas-phase and solid-phase abundances, respectively. We adopt a H column density of log N$_{\rm H}$ (cm$^{-2}$)$=$23.3 \cite{Pereira24b} (see Methods). Note that the HCN/H$_2$O abundance gas-phase observed ($\sim$0.38) is higher than that observed in these hot cores ($\sim$0.04) \cite{van_Dishoeck98}. Error bars indicate 1$\sigma$ uncertainties.} 
\label{abundance_plot}
\end{figure*}

\noindent
{{\bf(1) High-temperature gas-phase chemistry:}} Gas-phase hydrocarbon abundances are known to depend on temperature, as seen in massive star-forming regions \cite{Doty02}, AGB stars \cite{Agundez08}, and protoplanetary disks \cite{Bast13}. 

The kinetic temperature estimated in the molecular gas of IRAS\,07251$-$0248 probed by JWST is only moderately high, 150-250\,K. UV or X-ray photons are heavily attenuated in deeply obscured nuclei. Therefore, to evaluate whether standard gas-phase chemistry can account for the observed abundances of organic molecules, we used a chemical model without UV or X-ray photochemistry but including a high CR ionization rate, derived from the H$^+_3$ abundance in this source \cite{Pereira24b}.

As seen in left panel of Extended Data Fig.\,7, at the observed kinetic temperature and under a standard oxygen-rich scenario (C/O=0.55), the abundances of HCN, C$_2$H$_2$, CH$_4$, and the highly reactive CH$_3$ radical lie well below the observed values. While moderately high temperatures ($\sim$200\,K) do not account for the observed abundances, most species are significantly boosted above 500–600\,K, such that very hot chemistry occurring near the compact hot component, and then outflowing at $\sim$160\,km/s (see Methods) into the warm shell, cannot be ruled out. However, the hot chemistry model still underestimates some hydrocarbons, particularly CH$_4$ ($\sim$100 times more abundant than predicted; see Methods). Furthermore, chemical equilibrium is reached quickly, with a chemical relaxation time of $\sim$8$\times$10$^{2}$$\times$(10$^{5}$/$n_{\rm H}$\,cm$^{-3}$)\,yr, defined as the time required for the CH$_4$ abundance to reach a value within a factor of two of its steady-state abundance, with CH$_4$ being among the slowest species to reach the steady-state (see Methods; Extended Data Fig.~8). Therefore, standard high-temperature gas-phase chemistry cannot fully explain the observations.\\

\noindent
{{\bf(2) Oxygen depletion:}} Observations of protoplanetary disks around low-mass stars have indicated very high C/O ratios and rich hydrocarbon-chemistry \cite{Tabone23}. One proposed explanation is the depletion of oxygen caused by the formation of pebbles and planetesimals that trap water ice \cite{Tabone23}. This mechanism effectively raises the gas-phase C/O ratio, favoring the efficient formation of hydrocarbons. 

IRAS\,07251$-$0248 hosts a strong IR-emitting source, likely a deeply obscured AGN, as indicated by its IR luminosity surface density \cite{Pereira-Santaella21}. This AGN heats the surrounding gas, producing a warm shell ($\sim$150-250\,K) where the gas-phase molecules are observed. These temperatures are above the sublimation temperature of water ice, making this mechanism unlikely.\\

\noindent
{{\bf(3) Molecular ice desorption:}} It is well established that in very warm regions around protostars, sublimation of ice mantles drives an environment rich in organic molecules \cite{van_Dishoeck98,Herbst09}. While ice desorption is commonly considered to explain the presence of complex organic molecules (COMs; \cite{Molpeceres22}), it may also contribute to the abundance of small hydrocarbons \cite{Chuang21}.  CH$_4$ ices, for instance, have been detected in recent JWST observations of protostars and local IR galaxies \cite{Rocha24,Buiten25,Alonso-Herrero24}.

In IRAS\,07251$-$0248, a significant reservoir of gas-phase molecules and ices is present (see Fig. \ref{abundance_plot}). The molecular material forms a stratified structure \cite{Bernete24a}, with the warm shell ($\sim$150-250\,K), where the gas-phase molecular absorptions are produced, surrounding the nuclear dusty core, and a colder outer envelope where ices reside. The observed temperature of the gas-phase molecules in the warm shell is sufficient to sublimate all relevant ices (see also Scenario 2). 

However, the derived high CH$_4$/H$_2$O and HCN/H$_2$O gas-phase abundance ratios ($\sim$0.15 and $\sim$0.37, respectively) are significantly higher than typically found in the ices of Galactic star-forming regions (CH$_4$/H$_2$O$\sim$0.026 and HCN/H$_2$O$\lesssim$0.009; e.g. \cite{McClure23}). In addition, gas-grain chemical models of dense interstellar clouds predict ice abundance ratios of CH$_4$/H$_2$O$\sim$0.017, HCN/H$_2$O$\sim$0.054 and C$_2$H$_2$/H$_2$O$\sim$0.0006 at $\sim$3$\times$10$^5$\,yr \cite{Hasegawa93}, all of which are below the observed gas-phase abundance ratios. This indicates that the sublimation of these ices, while enriching the gas phase with hydrocarbons, cannot fully account for the observed hydrocarbon gas-phase abundances in the warm shell. Additionally, while not strictly considered ices, thermal desorption of carbonaceous grains may also contribute to the observed carbon enrichment in the gas phase (see Scenario 4).\\

{{\bf{(4) Carbon enrichment scenario: Carbonaceous grains and PAH erosion.}} In the standard oxygen-rich ISM, most carbon is locked in CO, limiting hydrocarbon formation. In contrast, environments with a higher C/O can maintain high abundances of simple organic molecules, such as C$_2$H$_2$, CH$_4$, and HCN \cite{Najita11}. Similar conditions are found in carbon-rich star envelopes \cite{Matsuura06} and M-type protoplanetary disks \cite{Tabone23}. However, unlike IRAS\,07251$-$0248, these sources generally do not exhibit a high abundance of H$_2$O and HCN. Moreover, deeply obscured ULIRGs differ markedly in global properties such as IR luminosity and dust mass, and, more importantly for the chemistry, exhibit higher densities and elevated temperatures compared to low-mass star-forming regions. 

Using the same chemical model described above, we examined whether a higher C/O ratio (C/O=1.30) could explain the observed gas-phase molecular abundances. At observed kinetic temperature ($\sim$200 K), the predicted C$_2$H$_2$ abundance approaches but remains below the observed value (Extended Data Fig. 7). However, CH$_4$, HCN and CH$_3$ are still largely underestimated by factors of $\sim$3.6$\times$10$^{3}$, 40, and 15, respectively. Only if we invoke a carbon-rich case combined with very high temperatures ($>$500-600\,K, which are higher than observed), as might be driven by the outflow, the observed values of some hydrocarbons are close to the model. However, the chemical relaxation timescales (see Scenario 1), together with the fact that the predicted CH$_4$ abundance is still $\sim$60 times lower than observed, argue against an interpretation exclusively based on the outflow-driven hydrocarbon-rich chemistry. In any case, even if the carbon-rich model could explain some of the hydrocarbon abundances, an additional source of carbon is needed to increase the C/O ratio from a standard oxygen-rich scenario, because significant oxygen depletion is unlikely (Scenario 2).

A plausible source of carbon in general, and hydrocarbons in particular, is the desorption from carbonaceous dust grains and PAHs. Processing of standard dust grains (typical values of the Milky Way are silicate $\sim$53\% and graphite 47\%; e.g. \cite{Mathis77}) would generate a chemistry where a significant fraction of oxygen is also released. This picture does not support the observed gas-phase hydrocarbon-rich environment. Notably, amorphous carbon dust grains are generally more efficiently destroyed than silicates \cite{Nieva12}. Carbonaceous grains and PAHs are important carbon sinks and their grinding-down process seeds the ISM with a variety of carbon clusters, chains, rings and small hydrocarbons (e.g. \cite{Pety05,Tielens2021_book,Tabone23}). PAH processing can release small hydrocarbons through the loss of CH, C$_2$H$_2$, carbon chains, and other highly reactive radicals such as CH$_3$ \cite{Monfredini19,Tielens2021_book} that can contribute to the synthesis of gas-phase organic molecules. This scenario is in good qualitative correspondence with the fragmentation of PAH suggested by the PAH ratios observed in IRAS\,07251$-$0248. 

Given the solar carbon abundance (2.4-3.0$\times$10$^{-4}$; \cite{Asplund09}) and estimates of carbon locked in PAHs ($\sim$10\%; \cite{Tielens2021_book}), even their complete destruction in the warm shell would be insufficient (abundance of carbon in PAHs $\sim$2.7$\times$10$^{-5}$) to explain the observed hydrocarbon budget in IRAS\,07251-0248 (see Fig. \ref{abundance_plot}). Assuming solar carbon abundance, an additional carbon source might be necessary, such as the processing of carbonaceous dust grains (\cite{Alata15, Murga23, Tabone23}), which contain $\gtrsim$30\% of cosmic carbon \cite{Draine03}. This process can release species such as C$_2$H$_2$, CH$_4$ and carbon chains, the most plausible major contributors to the rich gas-phase hydrocarbon chemistry. Laboratory experiments by \cite{Dartois17} yielded a C$_2$H$_2$/CH$_4$ gas-phase abundance ratio ($\sim$3.5) released from carbonaceous grain erosion similar to that observed in the gas-phase of IRAS\,07251-0248 ($\sim$2.4).

\vspace{5.5pt}
\noindent
In what follows, we discuss possible mechanisms for the processing of dust grains and PAHs:

\vspace{3.5pt}
\noindent
\textbullet{} {{\textit{Thermal desorption.}} The deep carbonaceous absorption features observed are likely associated with the colder icy reservoir envelope. Thermal processing of carbonaceous grains in the warm shell can enhance hydrocarbon abundances by releasing CH groups into the gas phase, thereby increasing the C/O ratio. This injection of hydrocarbons can be supported by laboratory data showing that carbonaceous grains initiate the lose of hydrogen and CH groups at temperatures $\gtrsim$200\,K (comparable to the warm gas-phase shell), although the process becomes efficient only at $\gtrsim$600\,K \cite{Duley96}. Thermal desorption induced by shocks might be plausible. However, it would imply a non-local origin for the observed abundances, which is difficult to reconcile with the high column densities of highly reactive species such as CH$_3$ (see also Scenario 1).

The high abundances of hydrogenated amorphous carbons suggest a potential formation mechanism in the colder envelope, where outflowing molecular material may transport these molecules from the inner warm regions to the outer colder layers. As hydrocarbon fragments might attach to carbonaceous dust grains and gas-phase molecules freeze out, the outflowing material in these outer regions may lead to high abundances of H$_2$O ice and hydrogenated carbonaceous grains. Supporting this scenario, all the gas-phase molecules studied in this work are observed outflowing, with a blueshifted velocity of $\sim$160$\pm$35\,km/s.

\vspace{3.5pt}
\noindent
\textbullet{} {{\textit{Non-thermal desorption:}}
Sputtering is an important mechanism for dust erosion \citep{Sonnentrucker07}. In the extremely dense and shielded environments of buried nuclei (e.g. IRAS\,07251$-$0248), carbonaceous grains and PAHs might be protected from UV and X-ray destruction. Processes such as shocks may also contribute to the hydrocarbon chemistry, however, the comparison of outflow velocities and the relative abundance of hydrocarbons in a sample of (U)LIRGs shows no clear correlation (see Extended Data Fig. 9). However, CRs can penetrate high column density molecular clouds and uniformly permeate the ISM. Energy deposition events produced by CRs can trigger grain processing via sputtering under conditions where thermal desorption may be inefficient \citep{Bringa04}. Recent laboratory experiments on carbonaceous grains have reported the production of hydrocarbon fragments by CR simulated irradiation \cite{Pelaez18,Pino19}. Similarly, CRs are considered significant drivers of PAH fragmentation and erosion through Coulomb explosions \cite{Micelotta10,Chabot20}, as has been proposed for PAH fragmentation in AGN outflows \cite{Bernete24c}.  

To explore the role of CRs in carbonaceous grains and PAHs erosion, we compare the relative abundance of C$_2$H$_2$, one of the primary products of carbonaceous grains and PAH fragmentation \cite{Kress10,Alata15}, as a function of the CR ionization rate, traced by the H$^+_3$ absorption equivalent width (EW) \cite{Pereira24b}. H$^+_3$ absorption is significant in buried nuclei, especially in IRAS\,07251$-$0248, which shows the deepest H$^+_3$ features implying the highest CR ionization rates in the sample studied in \cite{Pereira24b}. We use the ratio between the EW of the C$_2$H$_2$ (Q-branch of the $v_5$ mode at 13.7\,$\mu$m) and H$_2$O\,6.05\,$\mu$m (main contribution from the 2$_{12}$-1$_{01}$ line) absorptions as a proxy of the C$_2$H$_2$/H$_2$O abundance ratio, as function of the H$^+_3$ P(3,3) feature EW (at 4.35\,$\mu$m), which traces the cosmic ray ionization rate. In Figure \ref{ew_ratio_h3plus}, we use H$_2$O abundance as reference since it is relatively stable in chemical models (e.g. \cite{Doty02}).

\begin{figure}
\centering
\par{
\includegraphics[width=8.0cm]{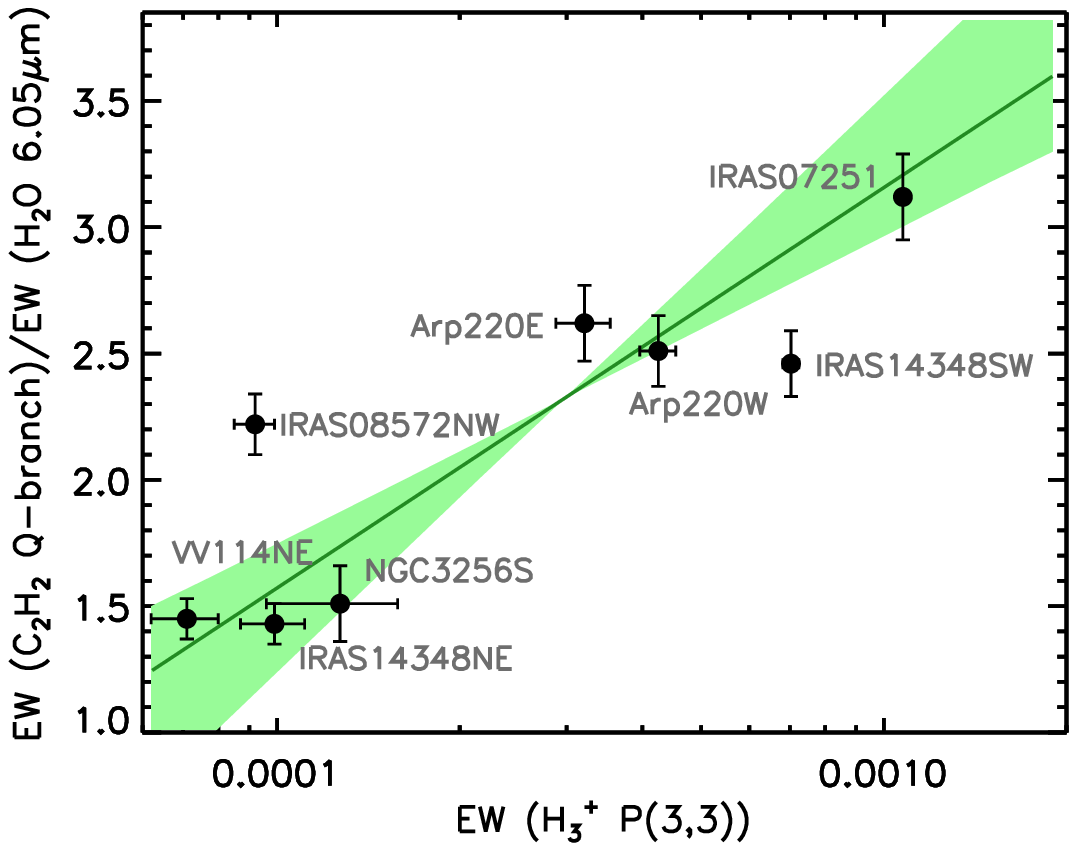}
\par}
\caption{Relationship between the C$_2$H$_2$/H$_2$O ratio and a cosmic-ray ionization tracer. Ratio between the equivalent width of the fundamental C$_2$H$_2$ (Q-branch at 13.7\,$\mu$m) and H$_2$O (at 6.05\,$\mu$m) bands versus the H$^+_3$ P\,(3,3) line (at 4.35\,$\mu$m), which is a good tracer of CR rate \cite{Pereira24b}. The green line corresponds to the best linear fit. To estimate the uncertainties in the fit (green area), we used bootstrap error estimation, generating 10$^{5}$ mock samples by randomly resampling the sources. Error bars indicate 1$\sigma$ uncertainties.}
\label{ew_ratio_h3plus}
\end{figure}

In Fig. \ref{ew_ratio_h3plus} we show the relative strength of the C$_2$H$_2$ to the H$_2$O absorptions as a function of the CR ionization for a sample of ULIRGs. We find a trend where high CR rates as traced by the EW of H$^+_3$ are associated with higher abundances of C$_2$H$_2$ relative to H$_2$O. While these results align with the proposed scenario to explain the rich hydrocarbon environment in IRAS\,07251$-$0248, future works involving detailed models of carbonaceus grain and PAH erosion by CRs would be valuable to further quantify their role in deeply obsucred AGN. For instance, although previous studies in the Galactic ISM have represented an important step forward, the CR energy distribution in AGN is expected to vary significantly (e.g. \cite{Koutsoumpou25}), and the properties of dust grains (e.g. size, porosity, structure) remain uncertain (e.g. \cite{Kirchschlager13,Ysard18}). 
\vspace{5.5pt}

Although turbulent mixing of gas at different temperatures is expected due to the outflow, the chemistry should quickly relax to local ($\sim$200 K) conditions, for which pure thermal desorption is not a plausible explanation. We conclude that, the observed C-rich chemistry characterizes outflowing gas presumably subject to post-shock conditions, but the dominant physical mechanism is most likely the processing of carbon grains and PAHs by CRs (see Extended Data Fig.~10).

\section*{Methods}\label{sec4}

\subsection*{Observations and data reduction}\label{subsec2}
IRAS\,07251$-$0248 (z$=$0.08775) was observed as part of the JWST GO Cycle 2 Large Program $\#$ID:\,3368 (P.I. L. Armus and A. Evans), for which near-IR to mid-IR (2.87-28.1~$\mu$m) integral field spectroscopy of the deeply obscured eastern nucleus was taken. In this work, we also include additional ULIRGs with H$^+_3$ P\,(3,3) detections from \cite{Pereira24b}, which are part of Program $\#$ID:\,3368 and the Guaranteed Time Observations Program $\#$ID:\,1267 (PI: D. Dicken). The data was observed using integral-field spectrographs MIRI MRS (4.9–28.1~$\mu$m; \cite{Wright15}) with a spectral resolution of R$\sim$3700--1300 and NIRSpec (\cite{Jakobsen22}) with the grating-filter pair G395H/F290LP (2.87–5.27~$\mu$m) with R$\sim$2700.

For the data reduction, we primarily followed the standard MRS pipeline procedure and the same configuration of the pipeline stages described in \cite{Bernete22d,Pereira22} to reduce the data. Some hot and cold pixels are not identified by the current pipeline, so we added an extra step before creating the data cubes to mask them as described in \cite{Pereira24,Bernete24a} for NIRSpec and MRS, respectively. We built a bad-pixel mask based on the background frames and leakage correction exposures and updated the data quality extensions of the observations \cite{Bernete22d,Pereira22}. The masked pixels are interpolated based on the neighbour spaxels when possible \cite{Bernete22d,Pereira22}.

\subsection*{Spectral extraction and continuum fit}\label{subsec3}
We extracted the central spectra by applying a 2D Gaussian model as described in \cite{Bernete24a,Bernete24b}. To do so, we employed observations of calibration point sources (HD-163466 and IRAS\,05248$-$7007, Programme IDs 1050 and 1049) to measure the width and position angle of a 2D Gaussian for each spectral channel. To obtain the point source flux we used the models of the calibration PSF stars from \cite{Bohlin20}, which is equivalent to applying aperture correction factors. This Gaussian defines the point spread function, which increases with wavelength. Keeping fixed the width and position angle values, we then fitted the Gaussian amplitude and position to the nuclear regions of source \cite{Bernete24a,Bernete24b,Pereira24}. In Fig. \ref{restframe_spec} we present the nuclear $\sim$2.9-28.1~$\mu$m spectrum of IRAS\,07251$-$0248. To examine the various molecular bands, we subtract a baseline representing the continuum emission from central spectrum. This is done by fitting feature-free continuum anchor points with straight lines to define the baseline as in \cite{Gonzalez-Alfonso24} and \cite{Bernete24b}. The fitted baseline is shown in Extended Data Fig. 1. 

\subsection*{Gas-phase molecule fit}\label{subsec4}
Synthetic spectra of the detected molecules were generated by following the same methodology as presented in \cite{Alfonso14,Gonzalez-Alfonso24, Pereira24,Bernete24b}, where a careful treatment of blending among lines of different species is considered. The level populations are assumed to be in LTE and the line profiles are reproduced by a Gaussian with the intrinsic velocity dispersion ($\sigma$) measured from non blended P- and R- branch lines of H$_2$O, HCN, and C$_2$H$_2$ ($\sim$105\,km/s), and the wavelength-dependent instrumental broadening from \cite{Jones23}. We constructed a grid of absorption models for IRAS\,07251$-$0248 covering a wide range of column densities, $T_{\rm kin}$, and relative abundances. To account for dilution effects on the observed absorption lines by additional mid-IR components in the extracted aperture, we introduce a covering factor (f$_{\rm cov}$= f$_{\rm obs}$ / f$_{\rm abs}$), where f$_{\rm obs}$ is the observed continuum (i.e., the fitted baseline) and f$_{\rm abs}$ is the absorbed continuum. The covering factor is defined as the fraction of the total observed continuum, including regions not intercepted by the molecular gas, relative to the continuum that produces the molecular absorption (f${\rm cov}$ = f${\rm obs}$ / f$_{\rm abs}$). This concept is equivalent to the background factor (f${\rm bg}$) defined in \cite{Buiten25}. For each synthetic spectrum, we computed the $\chi^2$ by comparing with the observed spectrum, while allowing the covering factor to vary. This $\chi^2$ was used to determine the best fitting parameters and uncertainties presented in Table~\ref{table_columns}. We note that the discrepancies between the HCN and C$_2$H$_2$ column densities derived from the different vibrational modes are related to their underlying continuum. At 7\,$\mu$m, the continuum is dominated by the compact hot component, whereas at 14\,$\mu$m additional contributions from cooler components dilute the emission of the compact hot component, leading to apparently lower column densities when using the long-wavelength bands. If the background continuum is not completely dominant, hidden line emission can partially fill in the absorption features. This effect is expected to be more significant at 14\,$\mu$m than at 7\,$\mu$m. Nevertheless, the high-J transitions in both the 7\,$\mu$m and 14\,$\mu$m bands trace the same gas. Such discrepancies have also been observed in other ULIRGs \cite{Buiten25}. We note that we measure an outflowing velocity of 160$\pm$35\,km/s relative to the rest-frame velocity of this object for all the gas-phase molecules studied in this work. The rest-frame velocity was determined from ALMA observations of the CO (2-1) transition \citep{Pereira-Santaella21}, and its uncertainty was estimated from the differences with the rest-frame velocity measured from stellar absorption features in the optical \cite{Perna21}. The analysis of additional molecular bands including other cyanopolyynes and cations will be presented in \cite{Speranza25}.

We used the spectroscopic parameters from the ExoMol database for HCN \cite{Barber2014}, H$_2$O \cite{Polyansky2018}, C$_2$H$_2$ \cite{Chubb2020}, CH$_4$ \cite{Yurchenko2024}, and CH$_3$ \cite{Adam2019}. For C$_4$H$_2$, we used the molecular parameters from the HITRAN database \cite{Gordon2022}. For C$_6$H$_2$ and C$_6$H$_6$, we derived the spectroscopic parameters from laboratory experiments and theoretical works.

\subsubsection*{C$_6$H$_2$}
The spectroscopic constants of the $v_{11}$ fundamental band of C$_6$H$_2$ were obtained from \cite{Haas1994v11} and the strength of this band from \cite{Shindo2003}. The energy of the $v_{13}$ mode of C$_6$H$_2$ is only 151\,K \cite{Haas1994v13}, so it can be significantly populated at low temperatures. The Q branches of the $v_{11}+v_{13}-v_{13}$ hot bands are blended with the fundamental $v_{11}$ band in the JWST spectrum. From laboratory experiments at $T=$250\,K, the hot band is expected to contribute up to $\sim$30\% of the absorbed flux (see Figure 3 in \cite{Haas1994v11}). Although we do not include the hot band in the model, we account for this uncertainty by adding a 30\% extra in the uncertainty of the derived column density.

\subsubsection*{C$_6$H$_6$}
For the $v_{4}$ fundamental band of benzene, we used the spectroscopic constants calculated by \cite{DangNhu1989} and the band strength from \cite{Sung2016}. The $v_{20}$ bending mode lies at a relatively low energy, 572\,K, and the $v_{4}+v_{20}-v_{20}$ hot band can contribute to the observed absorption at room temperature (the standard value often used laboratory experiments is 298\,K, corresponding to 25$^{\circ}$C; see Fig.~4 of \cite{Sung2016}). 

We included this hot band using the spectroscopic constants of $v_{20}$ from \cite{Pliva1993}. The spectroscopic constants and band origin of the $v_{4}+v_{20}$ mode were based on the $v_{20}$ constants, adjusted to match the position of the band and individual transitions measured by \cite{Kauppinen1980} and \cite{Sung2016}. The strength of this hot-band was estimated from \cite{Sung2016}.\\

\subsubsection*{Gas-phase H$_2$O fit}
Extended data Fig. 2 shows the best fitted model for the H$_2$O $v_2$=1-0 gas-phase molecular rovibrational bands of  IRAS\,07251$-$0248. %

\subsection*{Molecular solids band fits}\label{ices}

We examine dust absorption features and icy bands present in the mid-IR spectrum of IRAS\,07251$-$0248 (Fig. \ref{restframe_spec}). To do so, we computed the molecular solid strength (S$_{\rm solid~features}^{\lambda}$) following a method similar to that used by \cite{Bernete24a} and references therein, and we measured the ratio of observed flux (f$_{\rm obs}^{\lambda}$) to continuum flux (f$_{cont}^{\lambda}$) at the central wavelength of each silicate feature [S$_{\rm Sil}^{\lambda}=$ln(f$_{\rm obs}^{\lambda}/$f$_{\rm cont}^{\lambda}$)]. For the continuum curve, after masking broad PAH features and narrow emission lines, we assumed a spline (cubic polynomial) with feature-free continuum anchor points (see Extended Data Fig. 3). In this work, we refer to the ``optical depth curve'' ($\tau_\lambda$) as the molecular solid strength, calculated in the same way as for the silicate bands [i.e. $\tau_\lambda$=-ln(f$_{\rm obs}^{\lambda}/$f$_{\rm cont}^{\lambda}$), but using every wavelength ($\sim$5-30\,$\mu$m) \cite{Bernete24a}.

IRAS\,07251-0248 shows a very deep H$_2$O 6.0\,$\mu$m bending mode band ($\tau_{6\,\mu m}$$\sim$1.2) which is even deeper than that observed in NGC\,4418 by Spitzer (see Extended Data Fig. 3). While the H$_2$O libration mode absorption band around $\sim$11-16\,$\mu$m might be present \cite{Bernete24a,Bernete24b}, it is likely affected by blending with the broad 9.7\,$\mu$m silicate band ($\tau_{9.7\,\mu m}$$\sim$4). Thus, we focus on the 6\,$\mu$m bending mode, which is an excellent tracer of nuclear obscuration \cite{Bernete24a}. In contrast, the stretch mode at 3\,$\mu$m is mostly associated with the stellar continuum source (e.g. \cite{Rigopoulou24}) and might not be co-spatial with the material probed by the nuclear gas-phase molecular bands detected in the near- and mid-IR dust continuum. The H$_2$O 6.0\,$\mu$m bending mode, is generally mixed with those bands produced by a-C:H hydrogenated amorphous carbons (bands at $\sim$6.85 and 7.25\,$\mu$m; see Extended Data Fig. 4). {\textcolor{black}{In the solid phase, Methylene (CH$_2$) and CH$_3$ do not form isolated stable structures owing to their high reactivity, but are usually found in complex organic solids such as hydrogenated amorphous carbon grains (a-C:H; see e.g. \cite{Perrero23}).} The 6.85\,$\mu$m absorption band is generally attributed to the CH$_2$ and CH$_3$ bending mode while the less prominent 7.25\,$\mu$m absorption band is mainly associated with CH$_3$ \cite{Dartois07}. Furthermore, there might be some contribution of the icy band of CH$_4$ (see Extended Data Fig. 4), and we do not detect ices of C$_2$H$_2$ (strongest bands at $\sim$5.13, 7.19, and 12.35\,$\mu$m; \cite{Chuang24}) or HCN (strongest bands at $\sim$3.19 and 4.75\,$\mu$m; \cite{Gerakines21}). Using lab data, \cite{Gerakines21} identified the expected ice bands of HCN at $\sim$2.10, 2.37, 2.52, 3.19, 4.75, 6.16, and 12.12\,$\mu$m. 

When more than one ice species overlap in the same wavelength range, we decomposed the optical depth using lab data for each ice type. We used the H$_2$O and CH$_4$ ices from the Leiden Ice Database (LIDA, \cite{Rocha22}) together with a-C:H grains from \cite{Dartois07}. Decomposing the molecular absorption bands, we find $\tau^{\rm{H_2O}}_{6\mu m}$$\sim$1.2 and $\tau^{\rm{a-C:H}}_{6.85\mu m}$$\sim$1.5. We estimate the H$_2$O and CH$_4$ ice column densities by following \cite{Rocha22}, while CH$_{\rm n=2,3}$ column density of a-C:H grain's is calculated by using N(CH$_{\rm n=2,3}$)$\sim$2.5$\pm$0.7$\times$10$^{19}$\,$\tau$(6.85\,$\mu$m)\,cm$^{-2}$ from \cite{Dartois07}. {\textcolor{black}{The measured H$_2$O 6\,$\mu$m ice column density ($\sim$1.4$\times$10$^{19}$\,cm$^{-2}$) is higher than that measured for the gas phase ($\sim$5.4$\times$10$^{18}$\,cm$^{-2}$). We find a column density of N(CH$_{\rm n=2,3}$) $\sim$ 3.7-4.7$\times$10$^{19}$\,cm$^{-2}$ in a-C:H grains, even higher than the water ice column density (Fig. \ref{abundance_plot}). The derived a-C:H carbon-to-silicon ratio, [C$_{\rm a-C:H}$]/[Si]$\sim$5.2, estimated following the method described in \cite{Dartois07}, exceeds the average ($\sim$1.72) and maximum ($\sim$4.3) values found in a sample of local ULIRGs using Spitzer data \cite{Dartois07}, indicating a high column density of carbon locked in amorphous carbon dust. In contrast, the CH$_4$ ice band is almost undetected and the estimated column density is only an upper limit N(CH$_4^{\rm ice}$)$<$7.9$\times$10$^{16}$\,cm$^{-2}$.} To fit the CH$_4$ ice band, we first removed all fitted gas-phase molecular bands and then fitted the residuals. The right panel of Extended Data Fig. 4 shows the fitted CH$_4$ molecular solid band.

\subsection*{PAH modelling}\label{PAH_modelling}
To correctly investigate dust and PAHs, it is crucial to model the mid-IR spectra (e.g. \cite{Bernete24c,Donnan24,Rigopoulou24}). To do so, we use the differential extinction model by \cite{Donnan24}. In this model, the dust continuum, stellar continuum and PAH features are fitted simultaneously. This method implements a differential extinction model, where the continuum is the sum of a set modified blackbodies, with different temperatures and levels of extinction \cite{Donnan24}. This effectively probes different dust layers and provides satisfactory fits to deeply obscured sources. The PAH features are modelled as multiple Drude profiles, following previous methods such as PAHFIT \cite{Smith07a}. The extinction affecting PAH emission is generally lower than that inferred from the silicate absorption impacting the continuum. Therefore, we correct the PAH fluxes for extinction using H$_2$ lines, following the same approach as in \cite{Donnan24,Rigopoulou24}. In particular, we use the sensitivity of the H$_2$\,S(3) at 9.66\,$\mu$m to silicate absorption to estimate the extinction affecting this warm molecular gas. By adjusting the extinction correction to achieve a smooth rotation diagram, we infer the extinction value applied to the PAHs. We refer the reader to \cite{Donnan24} for a full description of the model. The fits for the 3.3, 3.4 and 11.3\,$\mu$m PAH bands are shown in Extended Data Fig. 5. In Extended Data Fig. 6, we show the 11.3/3.3\,$\mu$m PAH flux ratio as a function of UV radiation field hardness, traced by the [Ne\,III]\,15.55\,$\mu$m/[Ne\,II]\,12.81\,$\mu$m ratio \cite{Groves06}. We find no relation between the PAH ratios and the UV radiation field hardness.

\subsection*{Hydrogen column density}\label{subsec35}
For the warm shell containing the gas-phase molecules, we adopt a hydrogen nuclei column density of log N$_{\rm H}$ (cm$^{-2}$)$=$23.3, derived from the extinction affecting the mid-IR continuum \cite{Pereira24b, Donnan24}. This hydrogen column density is taken from \cite{Pereira24b} where the authors used the derived tau at mid-IR using the methodology described above to obtain the hydrogen column density by using an extinction law \cite{Bohlin78}.

\subsection*{Chemical model comparison}
\label{chem_model}

Here, we employed chemical models of dense obscured clouds, similar to those described in \cite{Agundez13}, to investigate whether the observed molecules can be accounted for by standard interstellar chemistry. These calculations (Extended Data Fig.~7) are based on a pseudo time-dependent chemical model including standard gas-phase processes. The model calculates the evolution of molecular abundances for fixed physical conditions (temperature, density, and CR ionization rate). We adopt the commonly used value for the H$_2$ formation rate coefficient of 3$\times$10$^{-17}$~cm$^{3}$~s$^{-1}$ \cite{Wakelam17}. We made use of a chemical network (list of processes and associated parameters) corresponding to the latest release of the UMIST Database for Astrochemistry \cite{Millar24}. At steady state, molecular abundances are sensitive to the ratio of the CR ionization rate of H$_2$, $\zeta$, to the volume density of H nuclei, $n_{\rm H}$, but not to $\zeta$ or $n_{\rm H}$ separately. In these models, we adopted a ratio $\zeta$/$n_{\rm H}$,\,=\,10$^{-18.2}$~cm$^{3}$~s$^{-1}$, as this value is inferred from H$_3^+$ observations of this source (\cite{Pereira24b,Speranza25}). Given the extremely obscured nature of this source (see Discussion), we included a high visual extinction (A$_{\rm V}$=30 mag) to prevent photochemistry due to external UV photons from taking place. We used roughly solar elemental abundances accounting for depletion of refractory elements on dust (table 3 of \cite{Agundez13}), which generally result in better agreement with the observations of dense clouds \cite{Agundez13}.  In this work, we focus on the chemical composition obtained after the system reaches steady state, which is achieved within 10$^2$-10$^3$\,yr for densities of 10$^5$-10$^6$\,cm$^{-3}$ (see Extended Data Fig.~8). The assumption of steady state is consistent with the quoted chemical timescales being much shorter than the dynamical time of $\sim$10$^5$ yr derived in {\textcolor{blue}{Gonz\'alez-Alfonso et al., in prep.}} Only at densities below $\sim$10$^3$\,cm$^{-3}$ which are unlikely in the source studied in this work, the chemical timescales would exceed $\sim$10$^5$ yr, in which case time-dependent effects need to be considered and the steady-state assumption might no longer be valid. The calculated steady-state abundances of several hydrocarbons are shown in Extended Data Fig.~7 as a function of the gas kinetic temperature. We carried out these calculations for the standard oxygen-rich (C/O$=$0.55; solid lines) and carbon-rich (C/O$=$1.30; dashed lines) scenarios. See \cite{Agundez13} for further discussion on chemical models.

Previous works found that some hydrocarbons may increase their abundance with temperature, with CH$_4$ showing the most significant enhancement at temperatures well above 400\,K \cite{Doty02}. However, these models assumed a $\zeta$/$n_{\rm H}$ ratio between $10^{-21}$ and $10^{-23}$~cm$^{3}$~s$^{-1}$, with the strong increase in hydrocarbons occurring at $\zeta/n_{\rm H} = 10^{-23}$~cm$^{3}$~s$^{-1}$ (their Fig. 4). Similar findings have been reported in other studies, for instance in the work of \cite{Agundez08}, which used a value of $\zeta/n_{\rm H} = 10^{-22}$~cm$^{3}$~s$^{-1}$. In addition, similar models have been explored in the context of AGN disks, assuming X-ray dominated regions (XDRs) \cite{Harada10}. 

However, the $\zeta/n_{\rm H}$ used in these studies are significantly lower than that observed in IRAS\,07251$-$0248, and a higher CR ionization rate might lead to more efficient destruction of molecules. Thus, in Extended Data Fig.~7, we examine the dependence of abundance on temperature using a representative value of the $\zeta$/$n_{\rm H}$ ratio for the source studied in this work. The abundance of all species in Extended Data Fig.~7 increases with temperature. However, Extended Data Fig.~7 also shows that the predicted hydrocarbon abundances relative to H are always lower than the observed ones at the observed kinetic temperature in IRAS\,07251$-$0248, even for the carbon-rich case. This suggests the need for an additional source of hydrocarbons in the nuclear region of this galaxy (see Discussion).

\bmhead{Data Availability}

The NIRSpec$+$MRS raw data used in this work is publicly available and can be accessed at the JWST archive (https://archive.stsci.edu/): proposal $\#$ID:\,3368 (P.I. L. Armus and A. Evans) and $\#$ID:\,1267 (PI: D. Dicken). The spectroscopic molecular parameters for all the species are available in ExoMol (\href{https://exomol.com/}{ExoMol}) and and HITRAN (\href{https://hitran.org/}{HITRAN}) databases, except for C$_6$H$_2$ and C$_6$H$_6$. The spectroscopic data for these two molecules will be shared on request to the corresponding author.

\bmhead{Acknowledgments}
The authors also acknowledge the DD-ERS, GTO and GO teams for developing the observing programs used in this work. The authors thank Aaron Evans, Bel\'en Mat\'e, Lee Armus, Patrick F. Roche, Ram\'on J. Pel\'aez and Susana Guadix-Montero for valuable discussions. IGB is supported by the Programa Atracci\'on de Talento Investigador ``C\'esar Nombela'' via grant 2023-T1/TEC-29030 funded by the Community of Madrid. IGB also acknowledges support from COST Action CA 21126 Carbon molecular structures in space (NanoSpace), supported by the European Cooperation in Science and Technology. MPS acknowledges support under grants RYC2021-033094-I and CNS2023-145506 funded by MCIN/AEI/10.13039/501100011033 and the European Union NextGenerationEU/PRTR. MPS and EG-A acknowledge funding support under grant PID2023-146667NB-I00 funded by the Spanish MCIN/AEI/10.13039/501100011033. EG-A thanks the Spanish MICINN for support under project PID2022-137779OB-C41. MA acknowledges funding support under grant PID2023-147545NB-I00. 

This work is based on observations made with the NASA/ESA/CSA James Webb Space Telescope. The data were obtained from the Mikulski Archive for Space Telescopes at the Space Telescope Science Institute, which is operated  by the Association of Universities for Research in Astronomy, Inc., under  NASA contract NAS 5-03127 for JWST; and from the European JWST archive (eJWST) operated by the ESAC Science Data Centre (ESDC) of the European Space Agency. These observations are associated with the programs \#1267 and \#3368. Finally, we thank the anonymous referees for their useful comments.

\bmhead{Author Contributions Statement} IGB, MPS and EGA contributed to the writing of the paper, methods and creation of radiative transfer models and figures. IGB, MPS, EGA, MA, DR, FRD, GS and NT have contributed to the interpretation of the results.

\bmhead{Competing Interests Statement}

The authors declare no competing interests. Correspondence and requests for materials should be addressed to IGB (email: igbernete@cab.inta-csic.es/igbernete@gmail.com).

\section{Extended Data}\label{secA1}
\renewcommand{\figurename}{Extended Data Figure} 
\setcounter{figure}{0}

\begin{figure*}[ht!]
\centering
\par{
\includegraphics[width=10.0cm, clip, trim=10 10 10 0]{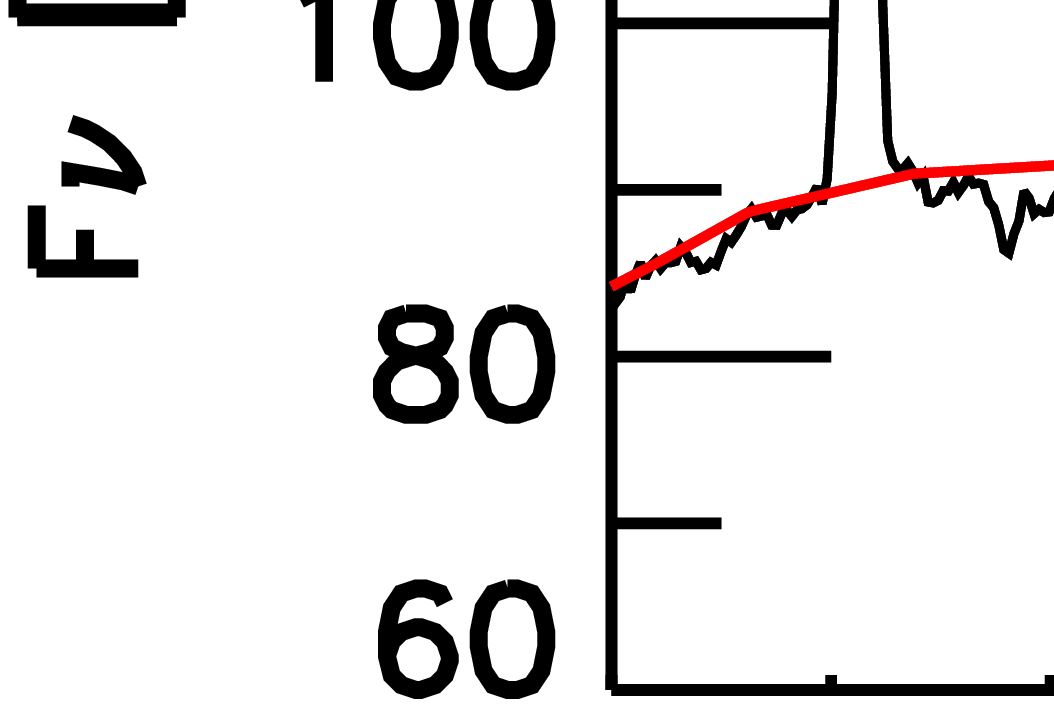}
\par}
\caption{Fitted baseline of IRAS\,07251$-$0248. The {\textit{JWST}} rest-frame spectra and baseline fit correspond to the solid black and red lines.}
\label{baseline}
\end{figure*}

\begin{figure*}[ht!]
\centering
\par{
\includegraphics[width=12.0cm, clip, trim=10 10 10 0]{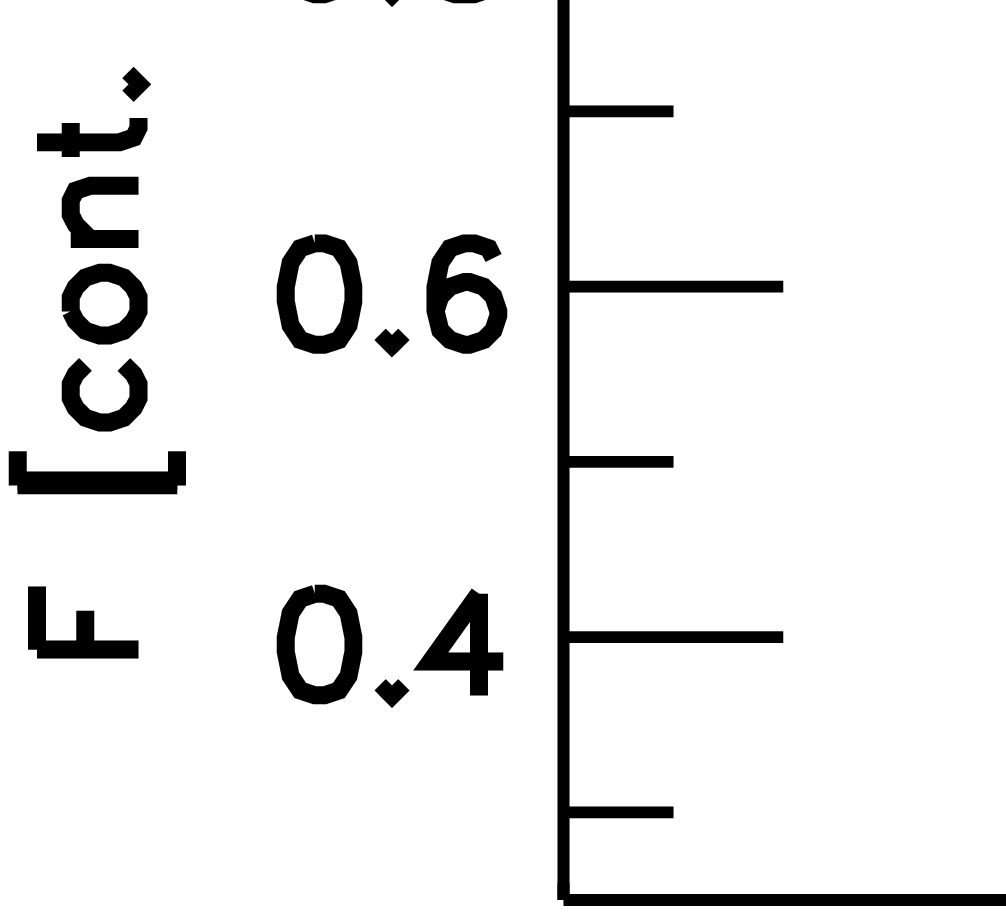}
\par}
\caption{H$_2$O $v_2$=1-0 (R-branch $\sim$5.3-6.2~$\mu$m) gas-phase molecular rovibrational bands in IRAS\,07251-0248. The {\textit{JWST}}/MIRI-MRS rest-frame continuum-normalized spectra (black solid line; filled in gray) is shown together with the total best-fit model (red solid line). The contribution of the fit (solid blue line) is shown with an offset ($+$0.15) to improve clarity.}
\label{model_water}
\end{figure*}

\begin{figure*}[ht!]
\centering
\par{
\includegraphics[width=6.5cm]{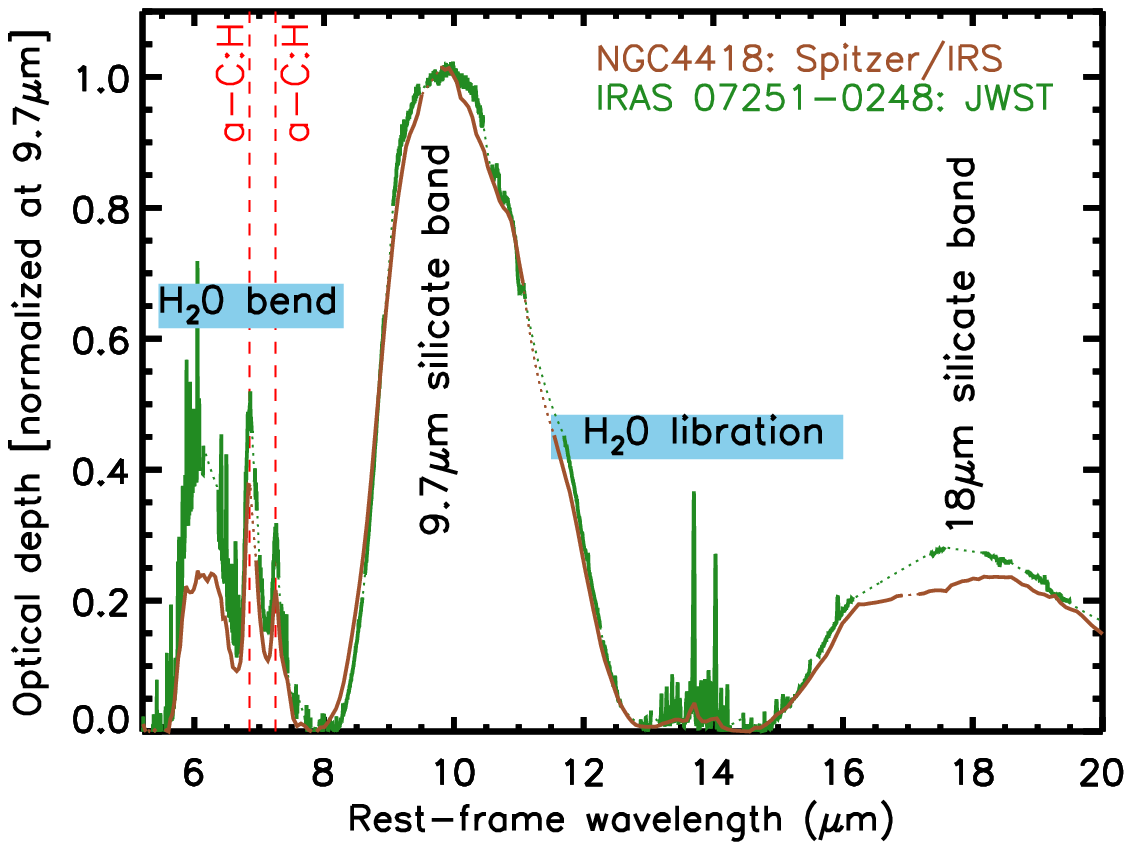}
\par}
\caption{Optical depth of IRAS\,07251$-$0248 normalized at 9.7\,$\mu$m. Dotted lines correspond to the masked spectral regions for removing the contribution of PAH features and narrow emission lines. Vertical red lines correspond to the main a-C:H hydrogenated amorphous carbon features at 6.85 and 7.25\,$\mu$m. The Spitzer/IRS spectrum of NGC\,4418, the prototypical CON, from \cite{Bernete24a}, is shown for comparison. \cite{Bernete24a} show a similar plot but using Compton Thick Seyfert galaxies.}
\label{opt_depth_appendix}
\end{figure*}

\begin{figure*}[ht!]
\centering
\par{
\vspace{-6pt}
\includegraphics[width=5.8cm]{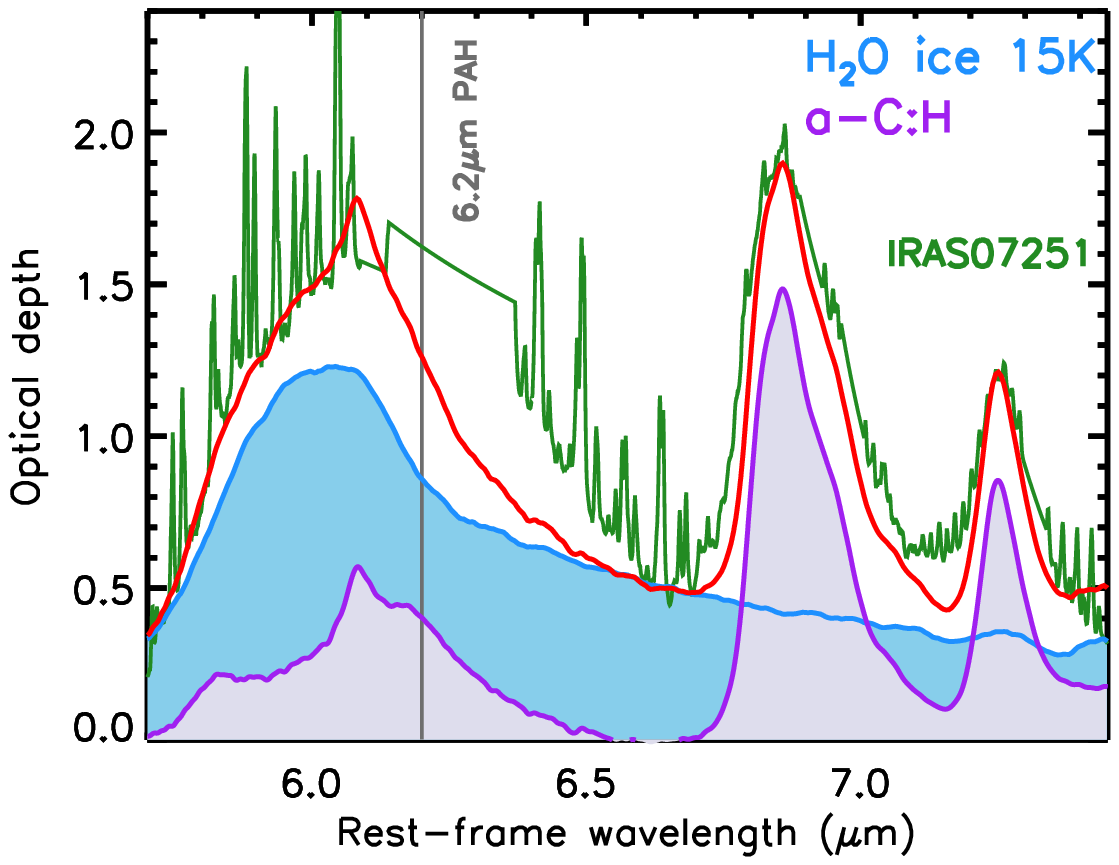}
\includegraphics[width=5.8cm]{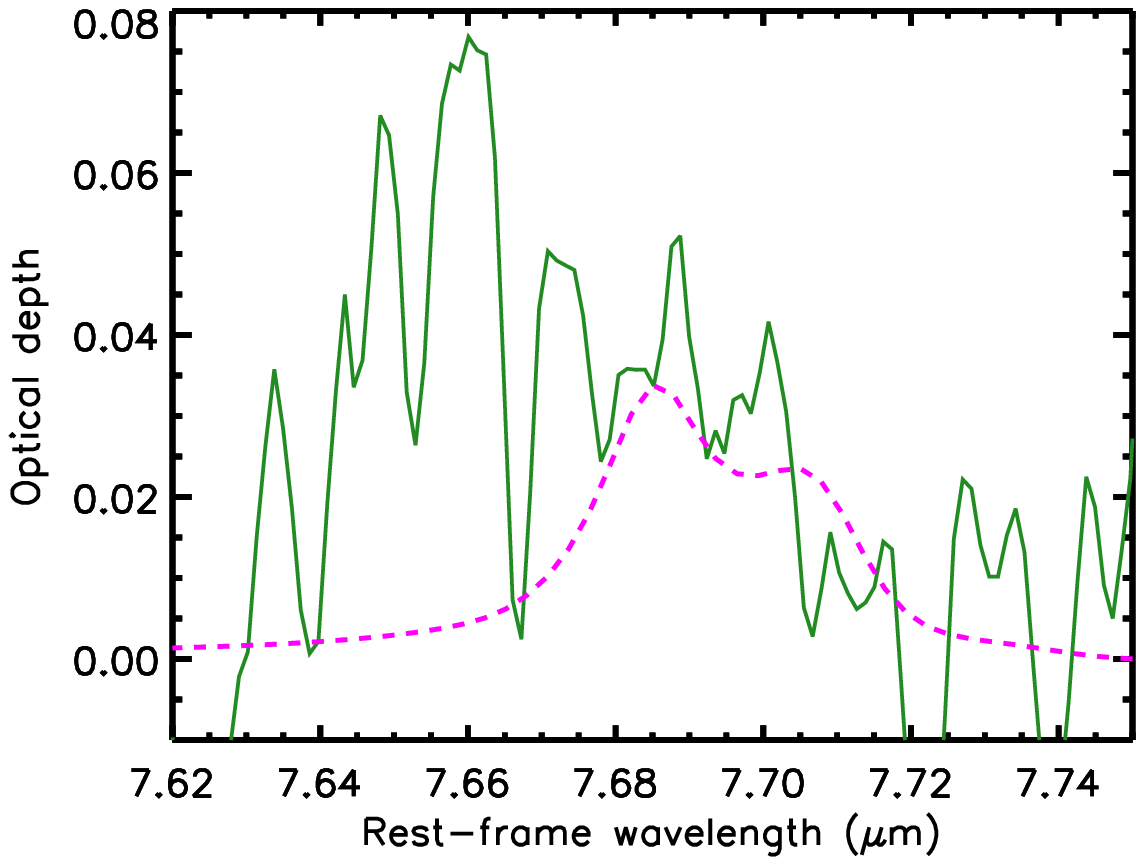}
\par}
\caption{Solid molecular absorption bands of IRAS\,07251-0248. Left: 5.5-7.5\,$\mu$m optical depth profile (green line). The best-fit model (solid red line) is a combination of laboratory spectra of pure water (blue shaded region corresponds to the H$_2$O bending mode at 15\,K; \cite{Ehrenfreund97,Rocha22}) and an a-C:H hydrogenated amorphous carbon analog (purple shaded region; \cite{Dartois07}). Right: 7.6-7.75\,$\mu$m optical depth profile (green line), which is fitted with the lab spectra of CH$_4$ ice profile from \cite{Rachid20} (magenta dashed line). Although the fit only approximates the observed features, it still provides an upper limit.}
\label{opt_depth}
\end{figure*}

\begin{figure*}[ht!]
\centering
\par{
\vspace{-12pt}
\includegraphics[width=5.7cm]{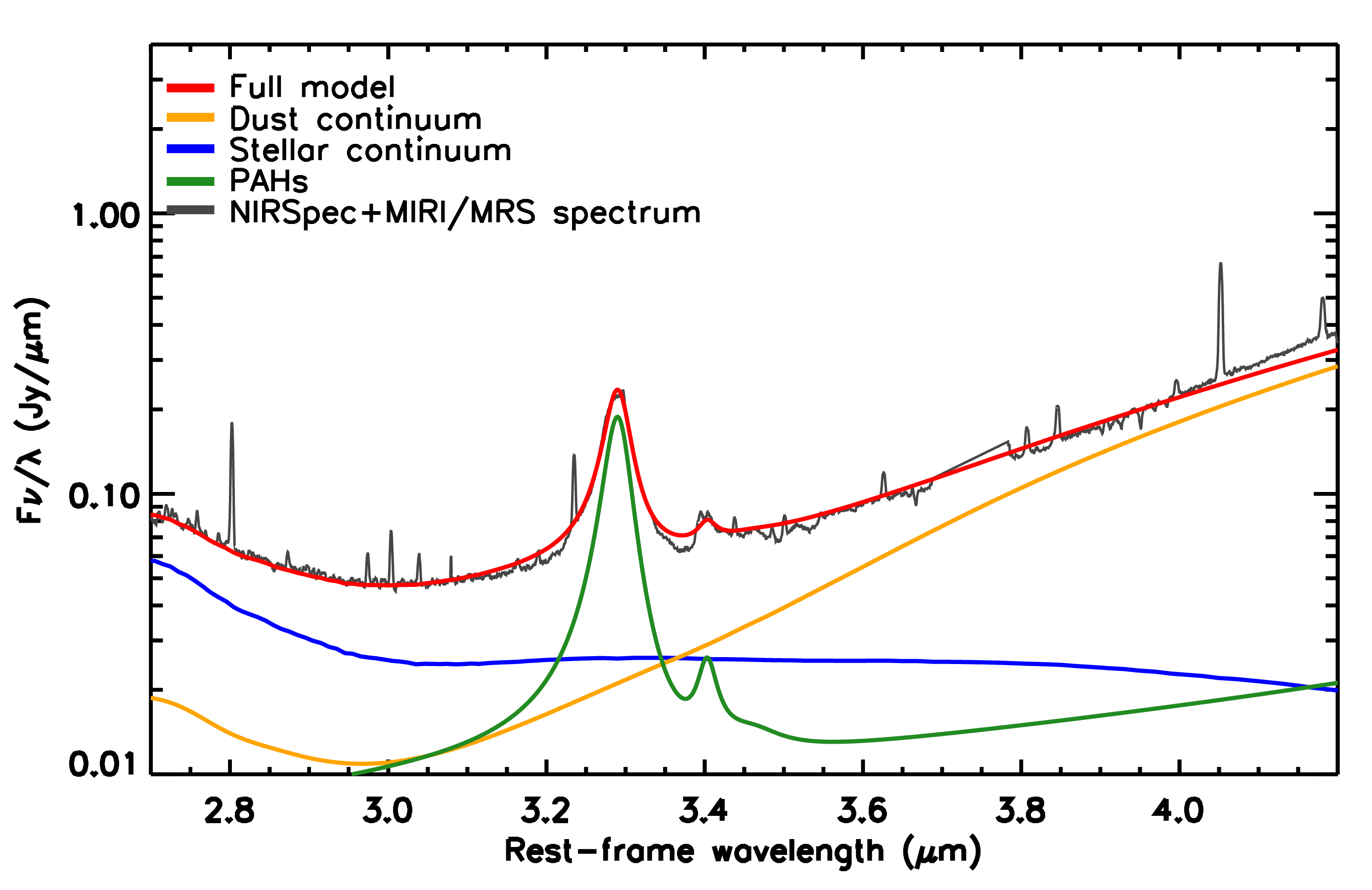}
\includegraphics[width=5.7cm]{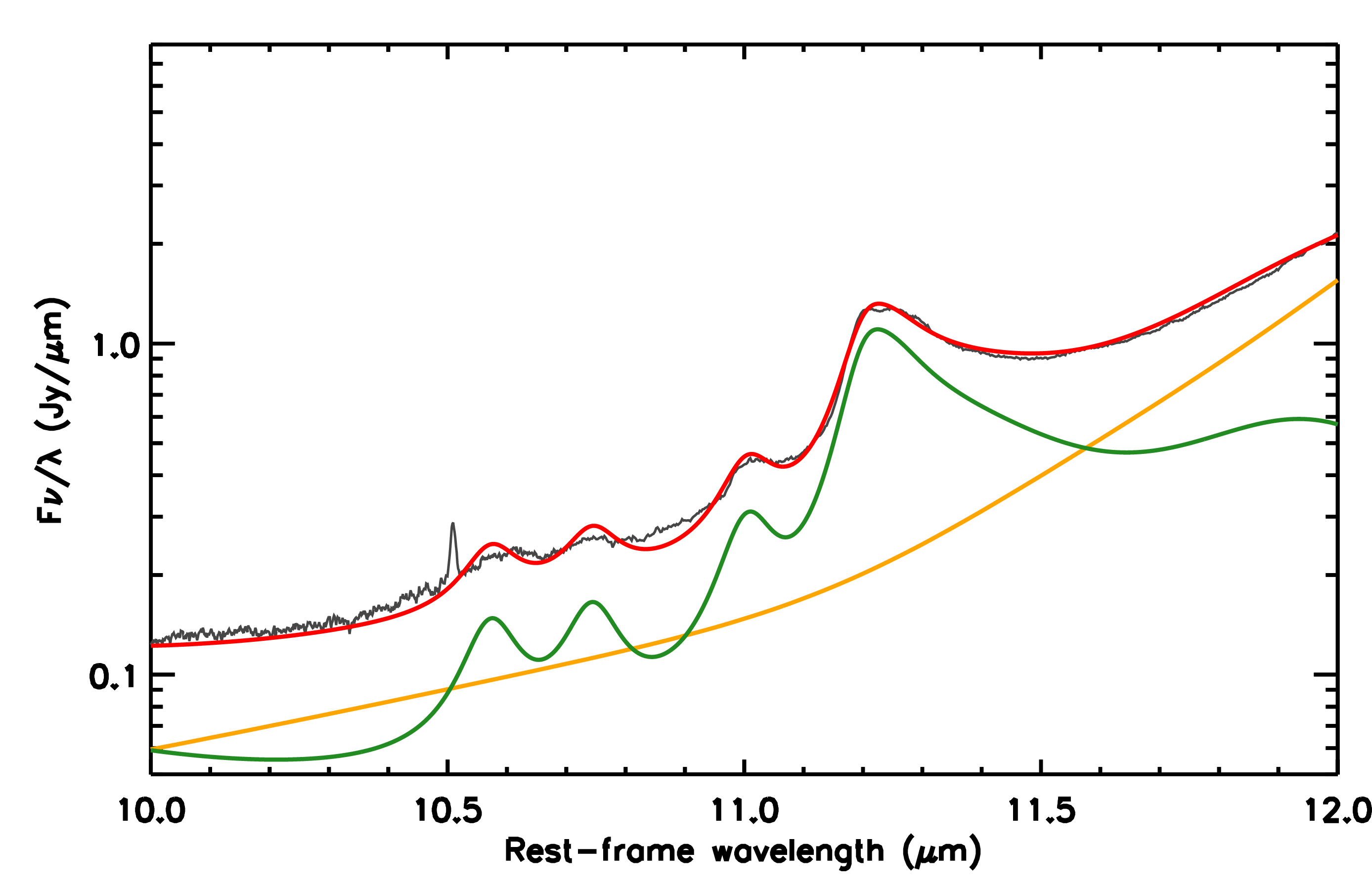}
\par}
\caption{Mid-IR spectral modelling of the nuclear region of IRAS\,07251-0248. Left panel: 
3.3, 3.4 and 3.47\,$\mu$m features. Right panel: 11.3\,$\mu$m PAH complex. The JWST rest-frame spectra and model fits correspond to the black and red solid lines. We show the dust continuum (orange solid lines), stellar continuum (blue solid lines) and the fitted PAH features (green solid lines).}
\label{pahplot}
\end{figure*}

\begin{figure*}[ht!]
\centering
\par{
\vspace{-10pt}
\includegraphics[width=6.0cm]{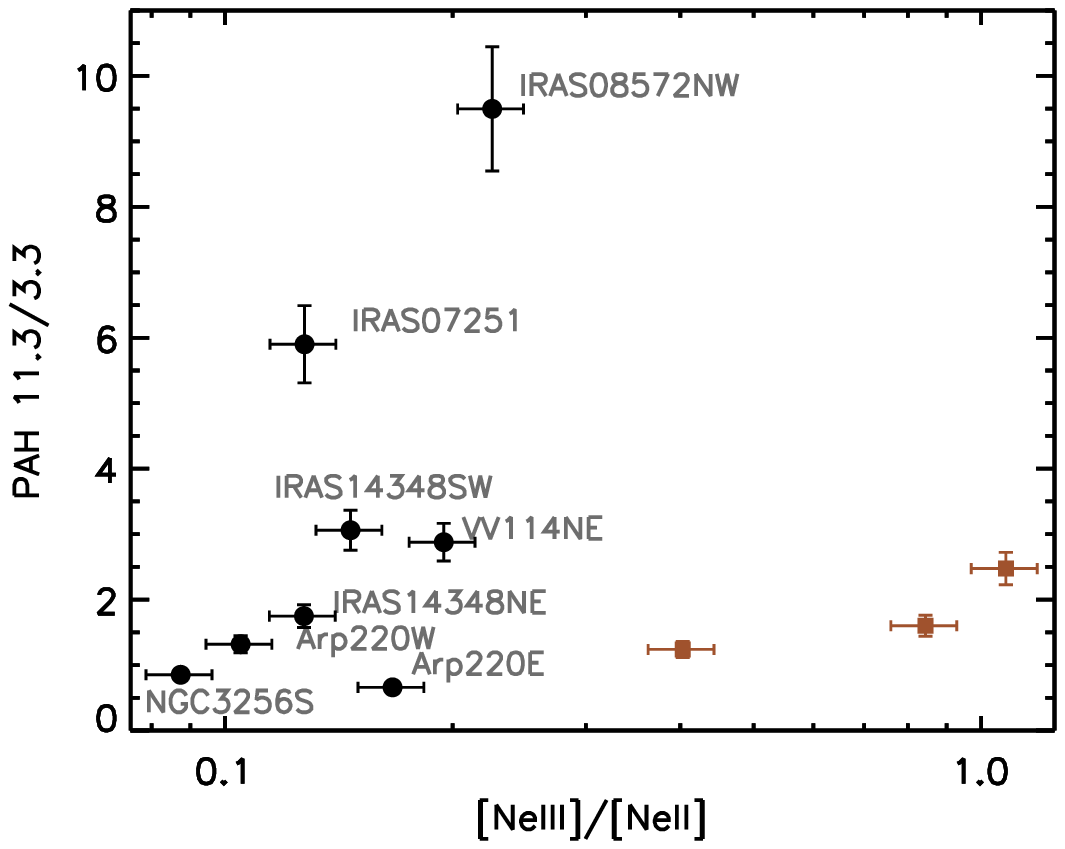}
\par}
\caption{11.3/3.3\,$\mu$m PAH flux ratio ratios vs. the hardness of the UV radiation field as probed by the [Ne\,III]\,15.55\,$\mu$m/[Ne\,II]\,12.81\,$\mu$m. The 11.3/3.3 PAH ratio indicates PAH molecular sizes
(e.g. \cite{Rigopoulou24}). The black circles represent the sources studied in this work. For comparison, we plot LIRGs (brown squares) from \cite{Rigopoulou24}. Error bars indicate 1$\sigma$ uncertainties.}
\label{hardness}
\end{figure*}

\begin{figure}[ht!]
\centering
\par{

\includegraphics[width=5.9cm]{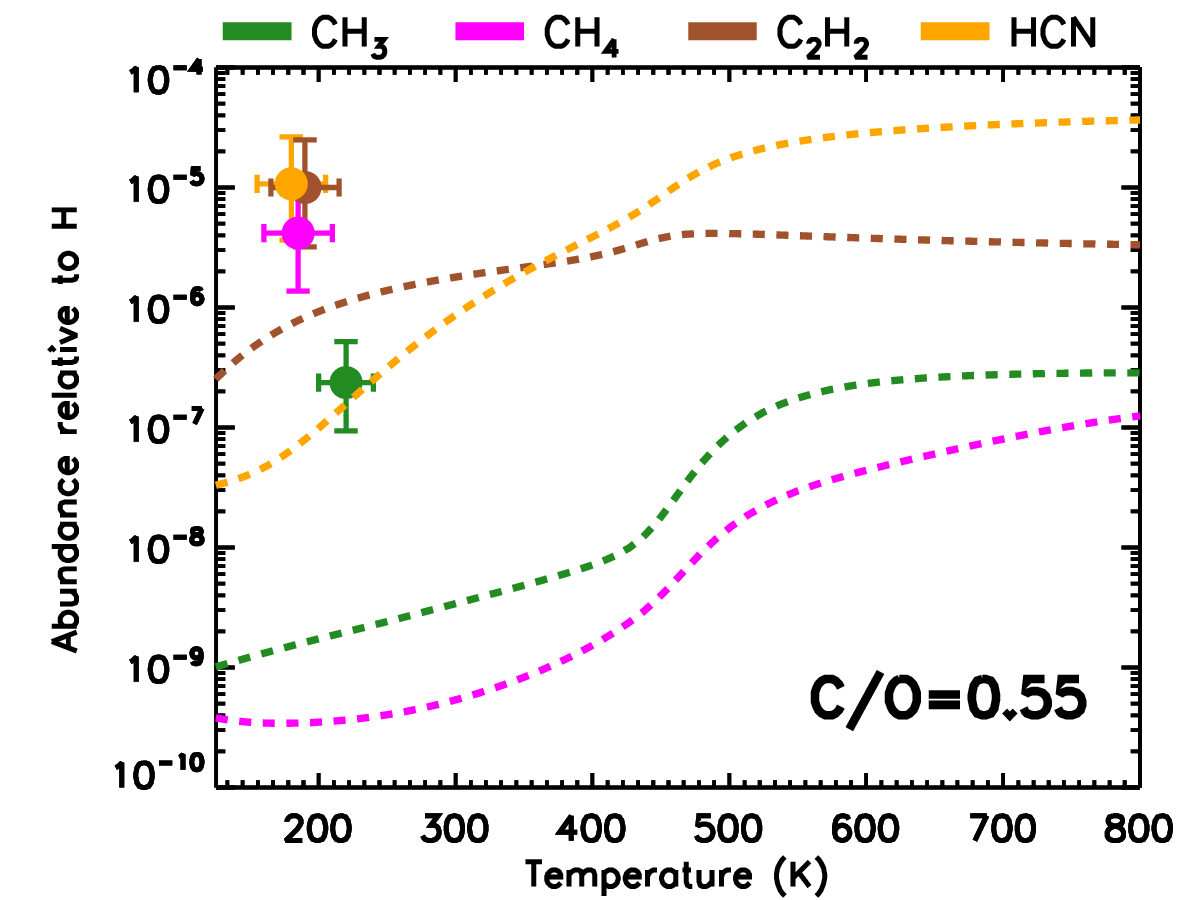}
\includegraphics[width=5.9cm]{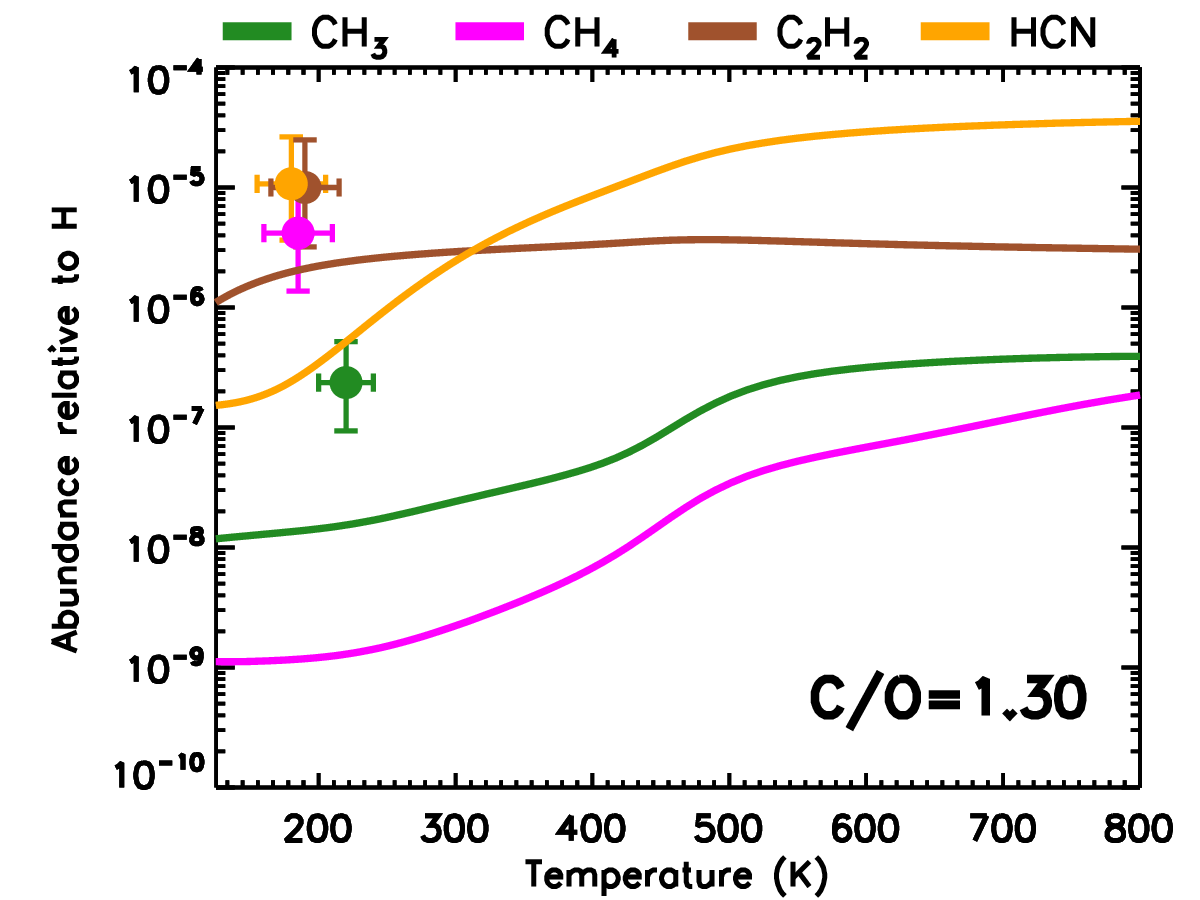}
\par}
\caption{Predicted fractional abundances of small organic molecules relative to H nuclei using a chemical model of a dense obscured cloud at steady state; see text and \cite{Agundez13} for a general view of the chemical model. Left panel:  standard oxygen-rich (C/O$=$0.55; dotted lines) scenario. Right panel: carbon-rich (C/O$=$1.30; solid lines) scenario. Circles correspond to the observed fractional abundance values for IRAS\,07251-0248. Note that HCN (orange circle) and C$_2$H$_2$ (brown circle) have practically the same abundance and kinetic temperature. Green, magenta, brown and orange lines correspond to CH$_3$, CH$_4$, C$_2$H$_2$ and HCN, respectively. The kinetic temperature of HCN and  C$_2$H$_2$ are shown with an offset ($\pm$10\,K) to improve clarity. Error bars indicate 1$\sigma$ uncertainties.} 
\label{model_abun}
\end{figure}

\begin{figure}[ht!]
\centering
\par{

\includegraphics[width=7.0cm]{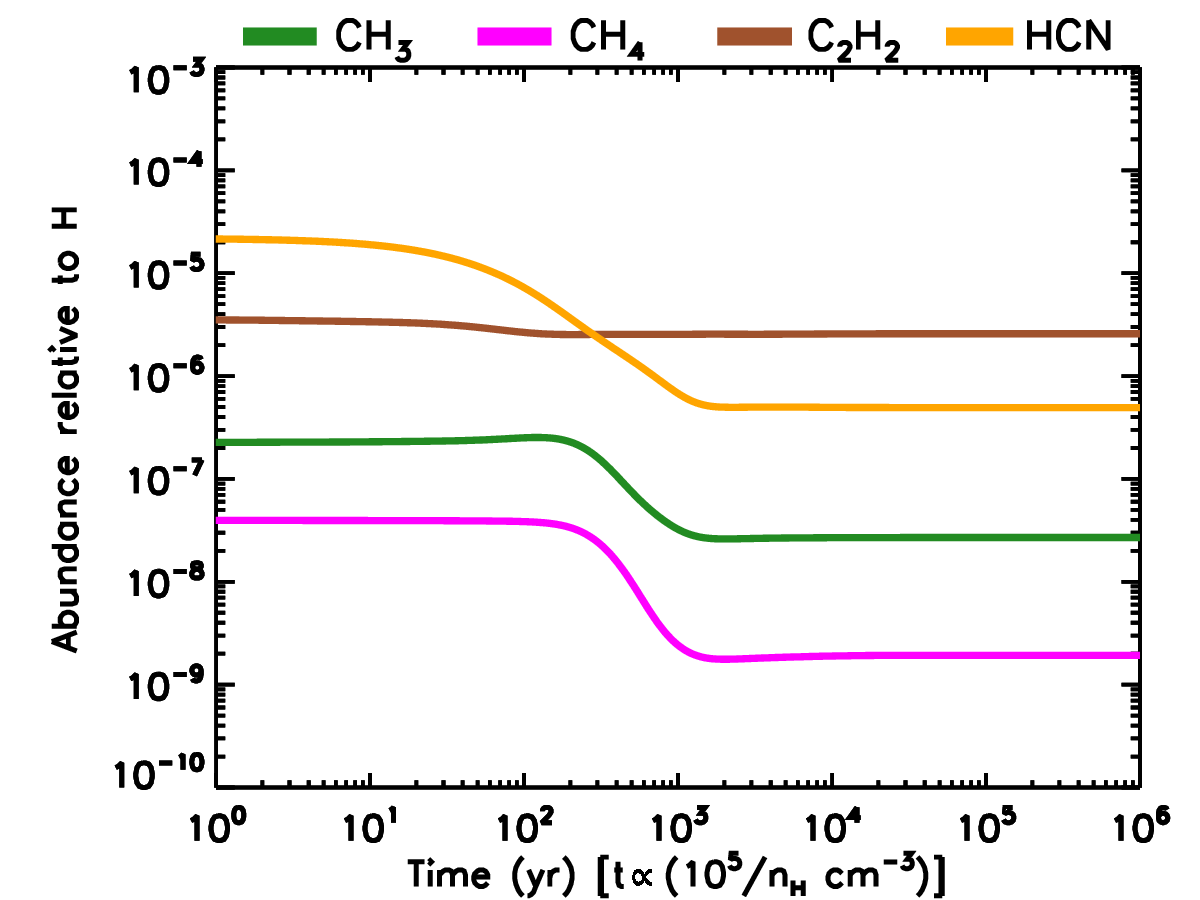}
\par}
\caption{Predicted chemical evolution of several organic molecules in the carbon-rich scenario (C/O$=$1.30). The initial conditions correspond to steady-state abundances at 500\,K; at t=0 the temperature is assumed to drop to 200\,K. Green, magenta, brown and orange lines correspond to CH$_3$, CH$_4$, C$_2$H$_2$ and HCN, respectively.} 
\label{relaxtime}
\end{figure}
\begin{figure*}[ht!]
\centering
\par{

\includegraphics[width=7.0cm]{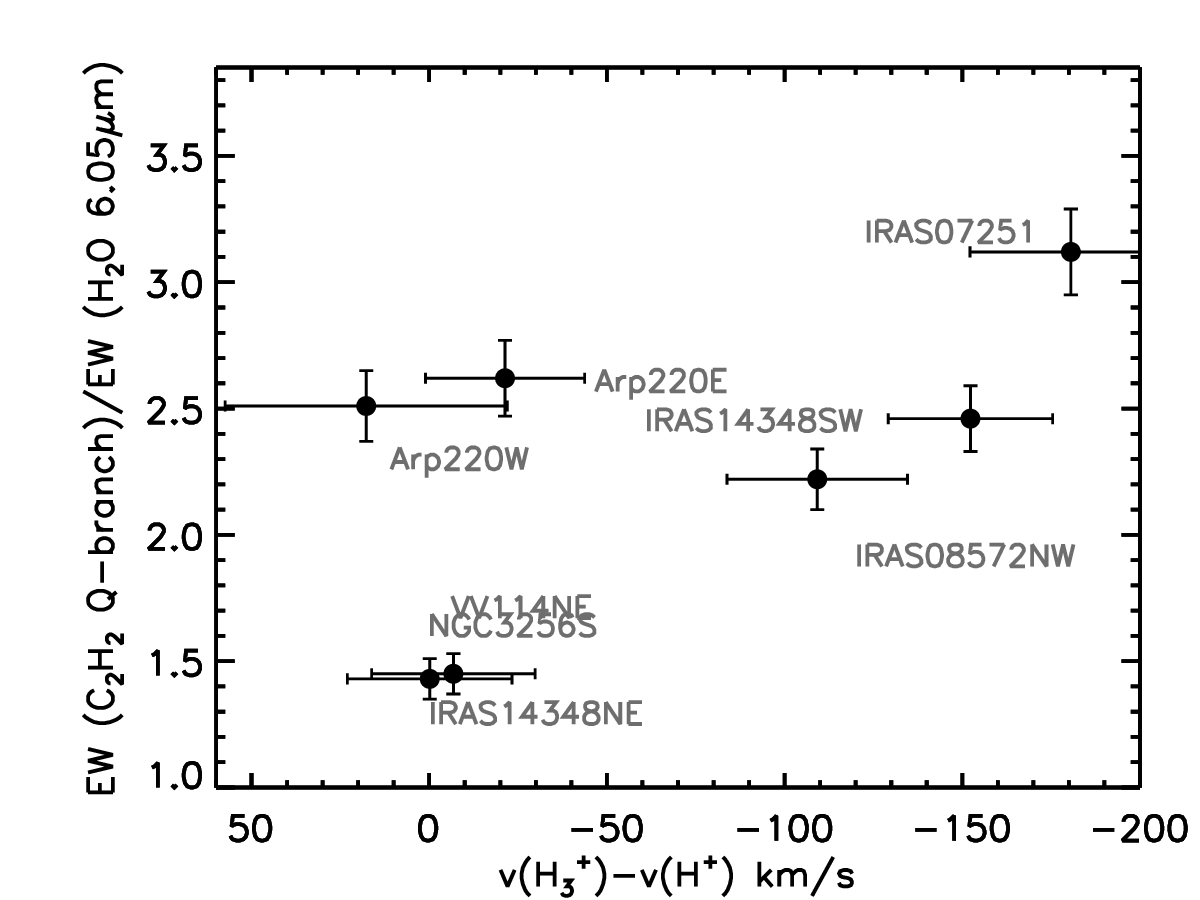}
\par}
\caption{Relationship between C$_2$H$_2$/H$_2$O ratio and a tracer of the velocity of the molecular outflow. Ratio between the equivalent width of the fundamental C$_2$H$_2$ (Q-branch at 13.7\,$\mu$m) and H$_2$O (at 6.05\,$\mu$m) bands versus the velocity measured for the H$^+_3$, which are from \cite{Pereira24b}. Error bars indicate 1$\sigma$ uncertainties.}
\label{shockss}
\end{figure*}

\begin{figure*}[ht!]
\centering
\par{
\includegraphics[width=11.6cm]{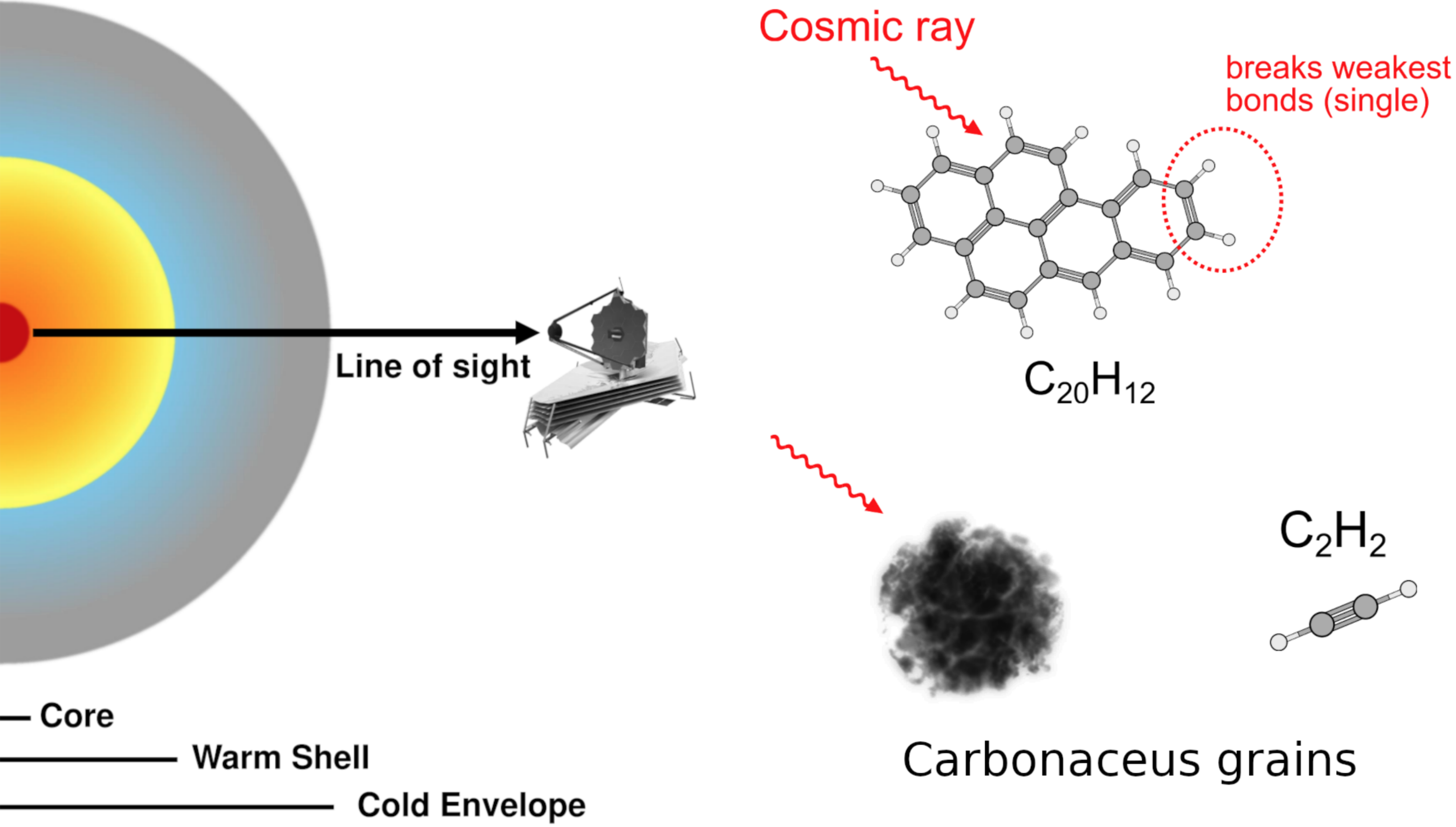}
\par}
\caption{Nuclear dust and molecular structure, with a sketch summarizing grain and PAH processing in IRAS\,07251$-$0248. Left panel: A schematic view of the central region of IRAS\,07251$-$0248. The color gradient represents decreasing temperature, from the extremely compact hot component (dark red; r$<$20\,pc), through the warm shell containing gas-phase molecules (orange-yellow; r$\sim$70\,pc {\textcolor{blue}{Gonz\'alez-Alfonso et al. in prep.}}), to the colder and icy envelope where solid-phase molecules reside (blue-gray). Right panel: Conceptual sketch illustrating the proposed scenario of carbonaceous grains and PAH processing driven by cosmic rays, which are responsible for hydrocarbon-rich chemistry. JWST icon credit: NASA Science Multimedia, \href{https://science.nasa.gov/wp-content/uploads/2023/05/webb_2.png}.}
\label{fig_abs}
\end{figure*}

\newpage
\textcolor{white}{.}
\newpage
\textcolor{white}{.}
\newpage
\textcolor{white}{.}
\newpage


\end{document}